\documentclass[useAMS,usenatbib]{mn2e}
\usepackage{amsfonts}
\usepackage{bbm}
\usepackage{graphicx}
\usepackage{times}
\usepackage{graphicx}
\usepackage{rotating}
\usepackage{array}
\usepackage{color}
\usepackage{epsfig}
\usepackage{amsmath}
\usepackage{natbib}

\title[Orbit classification in a N-body bar]
{Orbital classification in an N-body bar}
\author[Y. Wang et al.]
{Yougang Wang$^1$\thanks{E-mail:wangyg@bao.ac.cn},  E.~Athanassoula$^2$, Shude Mao$^{3,1,4}$  \\
$^1$Key Laboratory of Computational Astrophysics, National Astronomical Observatories,
Chinese Academy of Sciences, Beijing, 100012 China \\
$^2$Aix Marseille Universit\'e, CNRS, LAM (Laboratoire
d'Astrophysique de Marseille),
UMR 7326, 13388 Marseille 13, France\\
$^3$Center for Astrophysics, Department of Physics, Tsinghua University, Beijing, 100084, China\\
$^4$Jodrell Bank Centre for Astrophysics, University of Manchester, Manchester M13 9PL, UK }

\begin{document}

\date{Accepted . Received .}

\pagerange{\pageref{firstpage}--\pageref{lastpage}} \pubyear{2014}

\maketitle

\label{firstpage}

\begin{abstract}
The dynamics and evolution of any galactic structure are strongly
influenced by the properties of the orbits that constitute it.  In this paper, 
we compare two orbit classification schemes, one by Laskar (NAFF) ,
and the other by Carpintero and Aguilar (CA), by applying both of them to 
orbits obtained by following individual particles in a numerical
simulation of a barred galaxy.   We find that, at 
least for our case and some provisos, the main frequencies calculated by the
two methods are in good agreement: for $80\%$
of the orbits the difference between the results of the two methods
is less than $5\%$ for all three main frequencies. However,
it is difficult to evaluate  
the amount of regular or chaotic bar orbits in a given system. The
fraction of regular orbits obtained by the NAFF
method strongly depends on the critical frequency drift parameter,
while in the CA method the number of fundamental frequencies strongly
depends on the frequency difference parameter $L_{\rm r}$ and the maximum integer used for
searching the linear independence  of the fundamental frequencies. We
also find that, for a given particle, in general the projection of its
motion along the bar minor axis is 
more regular than the other two projections, while the
projection along the intermediate axis is the least regular.

\end{abstract}

\section{Introduction}
Nearly two-thirds of spiral galaxies in the Universe  have a bar
structure \citep[e.g.,][]{2012ApJ...745..125L, 2010ApJS..190..147B, 2015ApJS..217...32B}.
Bars are one of the main drivers for the secular evolution of disc galaxies (see \citealt{2013seg..book..305A} for a review), 
and can transport material from the bar region to the  center
and redistribute angular momentum within the
galaxy. This is emitted by the resonant regions in the bar and its vicinity, 
and absorbed by the outer parts of the disc 
and, mainly, by the spheroidal components (halo and bulge). Moreover, there
is a strong correlation between   
the strength of the bar and the amount of angular 
momentum thus redistributed 
\citep{2003MNRAS.341.1179A}. Therefore, understanding
the structure and the dynamical properties of bars is one of the most important issues in the formation and evolution of disc galaxies.

Orbits are the fundamental building blocks of any galactic structure
and therefore their properties greatly influence those  of the structure.
Moreover, it is difficult to describe the phase-space distribution for the chaotic orbits, which can not be adopted to construct torus models \citep{2008MNRAS.390..429M}.    
The orbit families and, more generally, the orbital structure in a fixed bar potential have been considered by many studies \citep{1980A&A....92...33C,1996MNRAS.283..149Z,2000MNRAS.314..433H,2011MNRAS.415..629M,2012MNRAS.427.1429W,2013MNRAS.435.3437W}.  Different methods of orbit classification have been used:

The Lyapunov exponent method \citep[see e.g. ][for a description]{1976PhRvA..14.2338B,1978CRASB.286..431B}. The Lyapunov exponents describe the time-averaged exponential rate of divergence of two orbits with close initial conditions in the phase space. Orbits with significantly non-zero Lyapunov exponents are chaotic.     
 
The Small ALignment Index method (SALI,  \citealt{2001JPhA...3410029S,2002MNRAS.337..619V,2004JPhA...37.6269S,2014MNRAS.438.2871C}). 
This method can be considered as an extension of the Lyapunov one, as
it relies on the properties of two arbitrary different initial
deviation vectors of an orbit, in order to distinguish efficiently between chaotic and regular orbits. The Generalized ALignment Index \citep[GALI,][]{2007PhyD..231...30S} is  similar to SALI, but uses a set of at least three initially linearly independent deviation vectors.
 
The NAFF method, short for Numerical Analysis of Fundamental
Frequencies, relies on the fact that the regular orbits move on a torus-like manifold and are thus quasi-periodic (\citealt{1990Icar...88..266L,1993PhyD...67..257L}). We will describe it further in Sect. 3.1.   

The spectral analysis method uses the Fourier transform of the time series of each coordinate of a given orbit  (\citealt{1998MNRAS.298....1C}, hereafter CA98). We will hereafter refer to this method as CA, from the initials of its authors, and describe it further in Sect. 3.2.  
 
 While each method has its advantages, each also suffers from
 disadvantages. For example,  the Lyapunov method necessitates very
 long integration times and the fraction of chaotic orbits also
 depends on the integration time ~\citep{1996ApJ...460..136M}; the
 SALI method also needs relatively long integration times, albeit much
 shorter than the Lyapunov method. The CA method has some problems for
 rotating systems \citep{2003CeMDA..85..247C} and depends strongly on
 the orbit integration time \citep{2012MNRAS.427.1429W}. Finally in
 the NAFF method whether an orbit is regular or not depends on the
 drift of its frequencies, so that a critical value needs to be
 adopted (See Sect. 5 in the present paper). Compared to other
 methods, CA and NAFF have an important advantage, namely they give
 more information for the regular orbits, such as their fundamental
 frequencies, from the ratios of which it is possible to define
 orbital families. Both of them have been successfully  applied to various potential systems  \citep[e.g.][]{1998A&A...329..451P,2010MNRAS.403..525V,2012MNRAS.422.1863B,2016ApJ...818..141V}.          

Most studies so far have relied on simple analytic potentials, which,
however, are not very realistic. 
In particular, real bars as well as N-body bars
are composed of two parts: an inner part which is thick both
horizontally and vertically, and an outer part which is thin in both
these directions, while as yet no analytical potential with such
a property has been developed (see \citealt{2016ASSL..418..391A} for a review). 
N-body bar potentials, however, are much more complex to use and
there are therefore relatively few studies relying on them, compared to
the large number of studies relying on analytic
potentials. \cite{2014MNRAS.438.2201M} and
\cite{2016MNRAS.458.3578M} took an intermediate path, using analytical
time-dependent potentials modelled after an N-body simulation of a
strongly barred galaxy. The disadvantage of this approach is that both the disc and the bar potentials are rigid and have not responded to each other,
which is not realistic.

An alternative route, much nearer to the N-bodies, is
to freeze the simulation potential at a representative time and then
follow in it orbits with initial conditions
obtained from the positions and velocities of the simulation particles
at that chosen time \citep{2002ApJ...569L..83A, 2003MNRAS.341.1179A,
2005NYASA1045..168A, 2006ApJ...637..214M, 2007MNRAS.381..757V, 2009A&A...494...11W,
2012MNRAS.419.1951V, 2016ApJ...818..141V}. This approach has a
number of advantages. The corresponding potentials are realistic, and
allow for orbital structure studies in bars with a thick inner part
and a thin outer part. It also provides a unique and correct
definition of the orbital sample which will be used, whereas in 
rigid potentials this sample is arbitrary, thus rendering any estimate of
the fraction of chaos in a given system also entirely
arbitrary. Indeed, whether a given orbit is  regular or chaotic
depends on its location within the galaxy's phase
space, and different samples may populate this space differently.
This severe drawback of analytical potentials is easily avoided by
relying on the simulation to provide the initial conditions of the orbits.
Concerning disadvantages, let us mention that a correct description of the
potential from the simulation particles is not trivial and also that
the potential has been frozen i.e. does not depend on time. It is
nevertheless possible to obtain information on time evolution by 
considering a series of consecutive times and of corresponding frozen
potentials. Thus full time information can be obtained, but in a very
time consuming manner.

A third alternative is to use directly the orbits of a preselected
number of particles during the simulation \citep{2007MNRAS.379.1155C,
2015arXiv151104253G, 2016arXiv160600322G}. This attractively straightforward way has a
number of difficulties, not the least being the fact that most of the
available techniques and information on orbital structure
have been obtained for non-evolving potentials. As we will show here,
however, this third alternative can still be very useful if one
chooses carefully the time interval over which one follows the 
orbits so that it has as little evolution as possible.

In this paper, we will give a detailed comparison of the CA and NAFF orbit classification methods
by studying orbits in a simulated bar. The outline of the paper is as follows. In \S 2 we describe briefly our numerically simulated bar.
In \S 3 we outline different methods of orbit classifications. In \S 4 we present the main frequencies from two methods. 
In \S 5 we present the fraction of  regular orbits from different classification schemes. In \S 6 we give a brief discussion. 
In \S 7 we present the summary and conclusions.       
      
\section{The simulation and bar orbits}

The initial conditions of this simulation comprise two components a
disc and a halo. Both are live, i.e. described self-consistently, in order
to allow exchange of angular momentum and thus a full bar growth
\citep{2002ApJ...569L..83A, 2003MNRAS.341.1179A}. The initial density
distribution of the disc is  

\begin{equation}
\rho_d (R, z) = \frac {M_d}{4 \pi h^2 z_0}~~\exp (- R/h)~~{\rm sech}^2 \bigg(\frac{z}{z_0}\bigg),
\end{equation}

\noindent
where $R$ is the cylindrical radius, $h$ is the disc radial scale length, $z_0$ is the disc vertical scale thickness and $M_d$ is the disc mass. 
The corresponding numerical values are $h=3$~kpc, $z_0=0.6$~kpc and
$M_{d}=5 \times 10^{10}~M_{\odot}$. For the halo we used an initial
volume density of 

$$
\rho_h (r) = \frac {M_h}{2\pi^{3/2}}~~ \frac{\alpha}{r_c} ~~\frac {\exp(-r^2/r_c^2)}{r^2+\gamma^2},
\nonumber\\
$$

\noindent
where $r$ is the radius, $M_h$ is the halo mass, $\gamma$
and $r_c$ are the halo core and cut-off radii, respectively, and the
constant $\alpha$ is given by 

$$
\alpha = \{1 - \sqrt{\pi}q~~\exp (q^2)~~[1-{\rm erf} (q)]\}^{-1}, 
$$
where $q=\gamma / r_c$ \citep{1993ApJS...86..389H}. The
numerical values used in this run are $r_{c}=42.4$~kpc, $\gamma=15$~kpc
and $M_{h}=19.54 \times 10^{10}~M_{\odot}$.     
The halo is described by 1 million particles and the disc has 200 000 particles. 

The initial conditions were built using the
iterative method of \cite{2009MNRAS.392..904R},
and to run the simulation we used a version of the GADGET3 code kindly
made available to us by V. Springel. For a full description of GADGET  
see \citep{2001NewA....6...79S, 2005MNRAS.364.1105S}. We adopted a softening
length of 100 pc for the disc and of 200 pc for the halo and an opening angle
of 0.5. 

With these initial conditions, the disc dominates the potential in the
inner parts, so that the bar forms very early on in the simulation.

\begin{figure}
 \centering
\includegraphics[width=80mm]{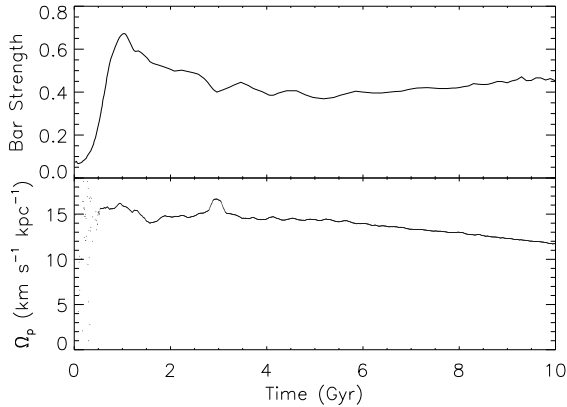}
\caption{Evolutions with time of the bar strength (top) and pattern speed (bottom) in our N-body bar. For more details, see \S 2. } 
\label{fig:bar_strength}
\end{figure}

\begin{figure}
 \centering
\includegraphics[angle=0, width=80mm]{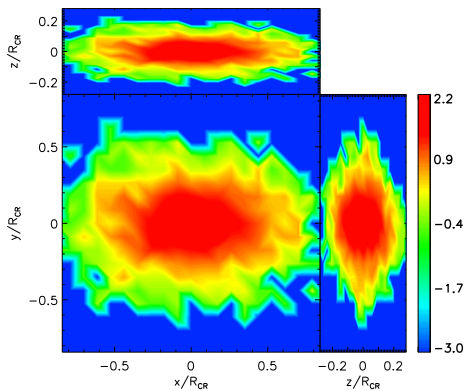}
\caption{Face-on (bottom left), side-on (upper left) and end-on (lower
  right) views of the distribution of the selected 3094 orbits at time
  6.0005 Gyr. The positions are
normalized by the corotation radius 
$R_{\rm CR}$.}
\label{fig:pos_ini}
\end{figure}

\begin{figure}
 \centering
\includegraphics[angle=0, width=80mm]{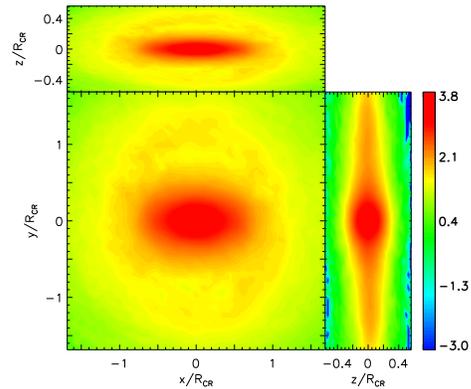}
\caption{Face-on (bottom left ), side-on (upper left) and end-on (lower
  right) views of the simulated disk at time
  6.005 Gyr.}
\label{fig:disk}
\end{figure}

The bar strength is defined as  in \cite{2013MNRAS.429.1949A}. More specifically,  
the Fourier components of the two-dimensional mass distribution can be written as 
\begin{equation}\label{eq:am}
a_m(R)=\sum_{i=0}^{N_R}m_i\cos(m\theta_i),\ m=0, 1, 2, ... 
\end{equation}  

\begin{equation}
b_m(R)=\sum_{i=0}^{N_R}m_i\sin(m\theta_i),\ m=1, 2, ... 
\end{equation}  
where $N_R$ is the number of the particles inside a given annulus around the cylindrical radius $R$, $m_i$ is the $i$th particle mass and 
$\theta_i$ is its azimuthal angle. The $a_m(R)$ and $b_m(R)$ are a function of the cylindrical radius. The bar strength is measured by the maximum amplitude of the relative $m=2$ component,
\begin{equation}
A_2=max\bigg( \frac{\sqrt{a_2^2+b_2^2}}{a_0}\bigg)
\end{equation}  
where $a_0$ is given by equation~\ref{eq:am} with $m=0$. The evolution of the bar strength and the pattern speed  with time are given in Fig.~\ref{fig:bar_strength}. We note
that in the time interval 6 to 10 Gyr the bar strength and the patter speed evolve
little with time, so we analyse the orbits in this time interval.
We selected a number of orbits visually, making
sure that they were in the bar at the time of selection (6 Gyr). We
then reran the simulation over the time range 6.0005-10.096 Gyr outputting only the positions, velocities and
accelerations of the selected particles, but for a very large number of times
(8192 outputs). We finally analysed 3094 orbits, whose initial positions at time 6.0005 Gyr  are shown in Figure~\ref{fig:pos_ini}.
The full disk at the nearby time (6.005 Gyr) is also presented in Figure~\ref{fig:disk}. It is seen that the disk has a more extended range than that of the selected orbits.  
Here and elsewhere in this paper, the positions of these orbits are
normalized by the corotation radius 
$R_{\rm CR}$.

\section{Orbit classification based on the Frequency maps }

The Fourier spectral analysis technique was pioneered by
\cite{1982ApJ...252..308B,1984MNRAS.206..159B} to classify regular and
chaotic orbits, and was then extended in different forms by
\cite{1993PhyD...67..257L} and CA98.  The key point of this method is
that regular orbits are quasi-periodic, thus the Fourier spectra
should consist of discrete lines and their frequencies can be
expressed as integer linear combinations of $N$ fundamental
frequencies (where $N$ is the dimension of the model). Chaotic orbits,
however, are  not quasi-periodic and the corresponding frequencies of
the Fourier spectra cannot be reduced to integer combinations of up to
only $N$ basic frequencies. 

Suppose that we have $N_d$ consecutive sampled values
$z_{k^\prime}\equiv z(t_k^\prime)$, where $t_k^\prime=k^\prime\eta$
, where $\eta$ is sampling interval, and $k^\prime=0,\ldots, N_d-1$. The discrete Fourier transform of 
$z_{k^\prime}$ can be written as      
\begin{equation}
Z_j=\frac{1}{N_d}\sum_{k^\prime=0}^{N_d-1}z_{k^\prime}\exp\bigg(-\frac{i2\pi jk^\prime}{N_d}\bigg),
\end{equation}        
where $j=-N_d/2+1,\ldots,N_d/2$. The Fourier spectrum consists of $N_d$ waves with amplitudes $|Z_j|$ and frequencies $\Omega_j=2\pi j/(N_d\eta)$.
We also define three amplitudes $|Z_{j,p}|$, $|Z_{j,v}|$, and  $|Z_{j,pv}|$: $|Z_{j,p}|$ and $|Z_{j,v}|$ correspond to the amplitudes from the position and velocity components, respectively, 
and $|Z_{j,pv}|$ is given by $\sqrt{|Z_{j,p}|^2+|Z_{j,v}|^2}/\sqrt{2}$. In this paper, we use $|Z_j|$ to represent $\sqrt{|Z_{j,p}|^2+|Z_{j,v}|^2}$ unless stated otherwise.
In order to facilitate the following discussions, we denote the time range 6.0005-8.048 Gyr as $t_1$, 8.0485-10.096 as $t_2$ and   6.0005-10.096 as $t_{\rm total}$ ($t_1=t_2=\frac{1}{2}t_{\rm total}$) .   

\subsection{NAFF}
The numerical analysis of fundamental frequencies (NAFF) was  pioneered by \cite{1990Icar...88..266L,1993PhyD...67..257L}, and developed further
by \cite{1996A&A...307..427P,1998A&A...329..451P} for both two and three dimensional models. The key point of NAFF is that regular orbits move on a torus-like manifold and are thus quasi-periodic. 

In an integrable system with $N$ degrees of freedom,  the Hamiltonian $H(\bf{J},\theta)$ depends only on the actions $J_j$, $H(J,\theta)=H(J_j)$, and the equations of motion of the system are given by
\begin{equation}
\dot{J_j}=0,~~~ \dot{\theta}_j(t)=\frac{\partial H}{\partial J_j}=\omega_j(\bf{J}),
\end{equation}
where  $\theta_j$ are angle variables, and $j=1, 2, . . ., N$. The orbit in the system can be written in terms of the complex variables  
\begin{equation}
z^\prime_j(t)=J_j e^{i\theta_j}=z^\prime_{j0}e^{i\omega_jt}
\end{equation}
where $z^\prime_{j0}=z^\prime_j(0)$. The motions in phase space take place on the surface of  tori that are products of  true circles with constant radii $J_j=|z^\prime_j(0)|$. The rate of  the motions around a torus is determined by the frequency vector $(\omega_1, \omega_2, ...,\omega_N)$. Generally, we do not know the precise action-angle variables $(J_j,\theta_j)$, but we can find approximations $(J^\prime_j,\theta^\prime_j)$. In the new coordinates, the motion can be written as
\begin{equation}\label{eq_ft}
f(t)=z^\prime_j(t)+\sum_{k}^{\infty}A_ke^{i\langle k,\omega\rangle t}  
\end{equation}  
where $A_k$ are the complex amplitudes, and $\langle k,\omega\rangle=k_1\omega_1+k_2\omega_2+...+k_N\omega_N$. In the limiting case, the coordinates $(J^\prime_j,\theta^\prime_j)$ are action-angle variables, and the amplitudes $A_k$ are close to zero.

In general, a system with more than one degree of freedom is not  integrable.  The Hamiltonian can be expressed as a perturbation of an integrable Hamiltonian $H_0$,
\begin{equation}
H(J,\theta)=H_0(J)+\epsilon H_1(J,\theta).
\end{equation}   
If the perturbation $\epsilon$ is small,  the Kolmogorov-Arnold-Moser (KAM) theorem suggests that a large fraction of the tori still exist and 
that the motion of most orbits is still quasi-periodic.   

The frequency map analysis consists of obtaining a quasi-periodic approximation of the numerical solutions of the Hamiltonian system in Eq.~(\ref{eq_ft}) in 
the form of a finite number of terms without searching for an explicit transformation of coordinates in action-angle variables    
\begin{equation}
f(t)=z^\prime_j(t)+\sum_{k=1}^{k_{\rm max}}A_ke^{i\langle k,\omega\rangle t}  
\end{equation}           
where $k_{\rm max}$ is the number of terms, and $A_k$ are of decreasing amplitude. 

A regular orbit is quasi-periodic, and the complex function combining its positions and velocities  $f(t)=X(t)+iV(t)$ can be expanded in a Fourier series \citep{2008gady.book.....B}
\begin{equation}
f(t)=\sum_{k=1}^{k_{\rm max}}A_k\exp(i\omega_kt)  
\end{equation}
where $\omega_k$ are the linear combinations of the fundamental frequencies , $\omega_k=l_k*\omega_1+m_k*\omega_2+n_k*\omega_3$, $A_k$ are the complex amplitudes and $k_{\rm max}$ is the number of terms.
The NAFF algorithm is designed to obtain an approximate form of $f(t)$     
\begin{equation}
f^{\prime}(t)=\sum_{k=1}^{k_{\rm max}}A^{\prime}_k\exp(i\omega^\prime_kt)  
\end{equation}    
where the frequencies $\omega^\prime_k$ and complex amplitudes $A^{\prime}_k$ can be obtained by an iterative scheme. The first frequency $\omega^{\prime}_1$ is searched by computing the maximum amplitude of 
$\phi(\sigma)=\langle f(t),\exp(i\sigma t)\rangle$ where the scalar product $\langle f(t),g(t)\rangle$ is given by
\begin{equation}
\langle f(t),g(t)\rangle=\frac{1}{T}\int_{-T/2}^{T/2}f(t)\bar{g}(t)\chi(t){\rm d}t,
\end{equation}    
where $T$ is the time interval, $\bar{g}(t)$ is the conjugate of $g(t)$, and $\chi(t)=1+\cos(2\pi t/T)$ is the Hanning window function. In the NAFF routine, the location of the primary frequency corresponds to the largest amplitude among the position spectrum  $|Z_{j,p}|$ and the velocity spectrum $|Z_{j,v}|$. The location of the first frequency is around the primary frequency. Once the first frequency has been found, its complex amplitude $A^{\prime}_1$ is obtained by  the orthogonal projection $A^{\prime}_1=\langle f(t),\exp(i\omega^\prime_1t)\rangle$. The first frequency component is subtracted and the process is restarted on the remaining part of the $f_1(t)=f(t)-A^{\prime}_1\exp(i\omega^\prime_1 t)$ to find the second frequency $\omega^\prime_2$. The process is repeated to find the third $\omega^\prime_3$, fourth $\omega^\prime_4$ and more frequency components until the residual function does not significantly decrease when subtracting the following term.  The fundamental frequencies are from these selected frequencies.             
                        
For the regular orbits, the fundamental frequencies do not change with time. Therefore, the frequency drift of the fundamental frequencies in two intervals provides us the regular behavior of the orbits . The frequency
drift is defined as  \citep{2010MNRAS.403..525V,2012MNRAS.419.1951V,2016ApJ...818..141V} 
\begin{equation}
\log(\Delta f_1)=\log\bigg|\frac{\omega_1(t_1)-\omega_1(t_2)}{\omega_1(t_1)} \bigg|,
\end{equation} 
\begin{equation}
\log(\Delta f_2)=\log\bigg|\frac{\omega_2(t_1)-\omega_2(t_2)}{\omega_2(t_1)} \bigg|,
\end{equation} 
\begin{equation}
\log(\Delta f_3)=\log\bigg|\frac{\omega_3(t_1)-\omega_3(t_2)}{\omega_3(t_1)} \bigg|,
\end{equation} 
and the frequency drift parameter  $\log (\Delta f)$  is  the largest value of    
$\log(\Delta f_1)$, $\log(\Delta f_2)$ and $\log(\Delta f_3)$. The orbit will be chaotic if the frequency drift parameter is large. 
Usually, a  critical value $\log(\Delta f_0)$ is used to distinguish chaotic from regular orbits. If the frequency drift is smaller
than the critical value $\log(\Delta f_0)$, the orbit is classified as regular, otherwise, the orbit is  chaotic.    
It is seen that the frequency drift in this definition is a relative drift; a shortcoming of this definition occurs when the fundamental frequency is large.
In particular, the accuracy of the determination of the main frequencies of the  ordinary FFT is of the order of $1/T$, and 
the NAFF method uses a Hanning  window to search for the maximum peak in the spectrum, which increases the accuracy of the 
main frequencies to the order of $1/T^4$ \citep{1996A&A...307..427P}. Thus, the frequencies of the orbits
can be recovered with high accuracy even for the chaotic orbits ~\citep{1998ApJ...506..686V}. If the absolute values of the fundamental frequency in the first and second intervals
are large, the relative value of  $|(\omega_i(t_1)-\omega_i(t_2))/{\omega_i(t_1)} |$ will still be small. Therefore, we also use a different definition of the frequency drift,
which is given by              

\begin{equation}
\Delta F_1=\bigg|\frac{\omega_1(t_1)-\omega_1(t_2)}{\delta \omega} \bigg|,
\end{equation} 
\begin{equation}
\Delta F_2=\bigg|\frac{\omega_2(t_1)-\omega_2(t_2)}{\delta \omega} \bigg|,
\end{equation} 
\begin{equation}
\Delta F_3=\bigg|\frac{\omega_3(t_1)-\omega_3(t_2)}{\delta \omega} \bigg|,
\end{equation}       
where $\delta \omega=2\pi/(N_d\eta)$, where $N_d$ is again the number of sampled points and $\eta$ is the sampling interval. The frequency drift parameter  $\Delta F$ is taken as the largest value of    
$\Delta F_1$, $\Delta F_2$ and $\Delta F_3$. We take this frequency drift as the absolute frequency drift. A critical value $\Delta F_0$ is used to distinguish regular orbits from  chaotic ones.           

\begin{figure}
\includegraphics[angle=0, width=80mm]{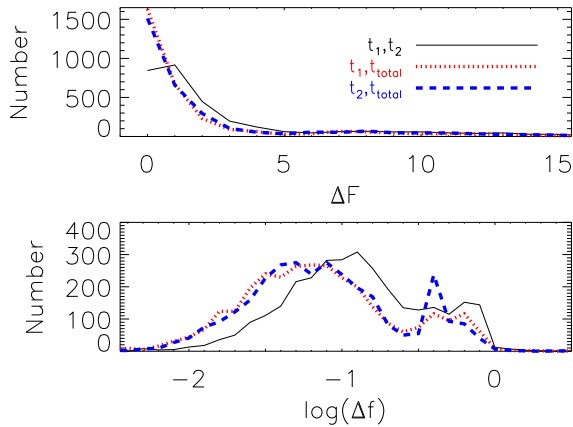}
\caption{Distribution of the absolute frequency drift parameter
  $\Delta F$ (top) and the relative frequency drift parameter
  $\log(\Delta f)$ (bottom) for our selected 3094 orbits. The solid, dotted and dashed lines 
   correspond to the comparison of different time ranges  as
  labelled at the top  
  right of the top panel.} \label{fig:df_dis}
\end{figure}

Figure~\ref{fig:df_dis} shows the distribution of
the absolute frequency drift parameter (top) and the relative
frequency drift parameter (bottom) from the NAFF method between $t_1$
and $t_2$ (solid line). It is seen that most orbits have an absolute frequency drift smaller than $2\delta \omega$. The peak of the distribution of the relative frequency drift $\log \Delta f $ is around -1, which indicates a $10\%$ frequency drift.

Generally, the frequency drift can be considered between  any two
different intervals, therefore, we also study the cases from the $t_1$
time range to $t_{\rm total}$ and $t_2$ to $t_{\rm total}$.  
In Figure~\ref{fig:df_dis}, we show the
frequency drift parameter from time $t_1$ to $t_{\rm total}$ (dotted lines), and one
from time $t_2$ to $t_{\rm total}$ (dashed lines), respectively. It is seen that most
orbits have smaller absolute frequency drift parameters in $t_1-t_{\rm
  total}$ and $t_2-t_{\rm total}$ than those in $t_1-t_2$, which can
be explained in the following way: The frequency resolution is twice
higher for time $t_{\rm total}$ time range than that for the $t_1$ and
$t_2$ ones since we use the same time step to output the orbits. The absolute frequency drift parameter is calculated  using $\delta \omega$ rather than $0.5 \delta \omega$ in cases  from time $t_1$ to $t_{\rm total}$ and $t_1$ to $t_{\rm total}$.  In this paper, the orbit types are given by using the drift parameter from time $t_1$ to $t_2$ unless stated otherwise.                

\subsection{The CA method}
The key point of CA is to find the number of the fundamental
frequencies. In its initial form this method  
used only the position to do the Fourier transform (CA98).  An updated version of
the code uses the Frequency Modified Fourier Transform
~\citep[FMFT]{1996CeMDA..65..137S} to extract lines,  and the spectral
analysis is performed on both the position and velocity component
$X(t)+iV(t)$, which is similar to what is done in the NAFF scheme.

In NAFF, the frequencies are calculated sequentially and any later frequency and amplitude depend on the previous ones.
Once the previous ones are found, they will not change in the
subsequent steps.
After $k-1$ cycles, the $k_{\rm th}$ frequency is shifted from
  $\omega^\prime_k$ mostly due to the existence of close
  frequencies which have significant amplitudes. After a number of cycles this can lead to differences of the order of several $\delta\omega$. The FMFT method consists of the NAFF process but gives a
  correction of frequencies via Eq. (36) in
  ~\cite{1996CeMDA..65..137S}. It is important to note that the
  frequencies and amplitudes in FMFT can change with the number of
  extracted lines because every frequency and amplitude are corrected
  by the primary selected peaks in the FFT spectrum. This is a major
  difference between the FMFT and the corresponding method used in
  NAFF (See ~Table\ref{table:1315peak} for an example).

\begin{figure*}
\begin{center}
\includegraphics[height=0.35\textwidth]{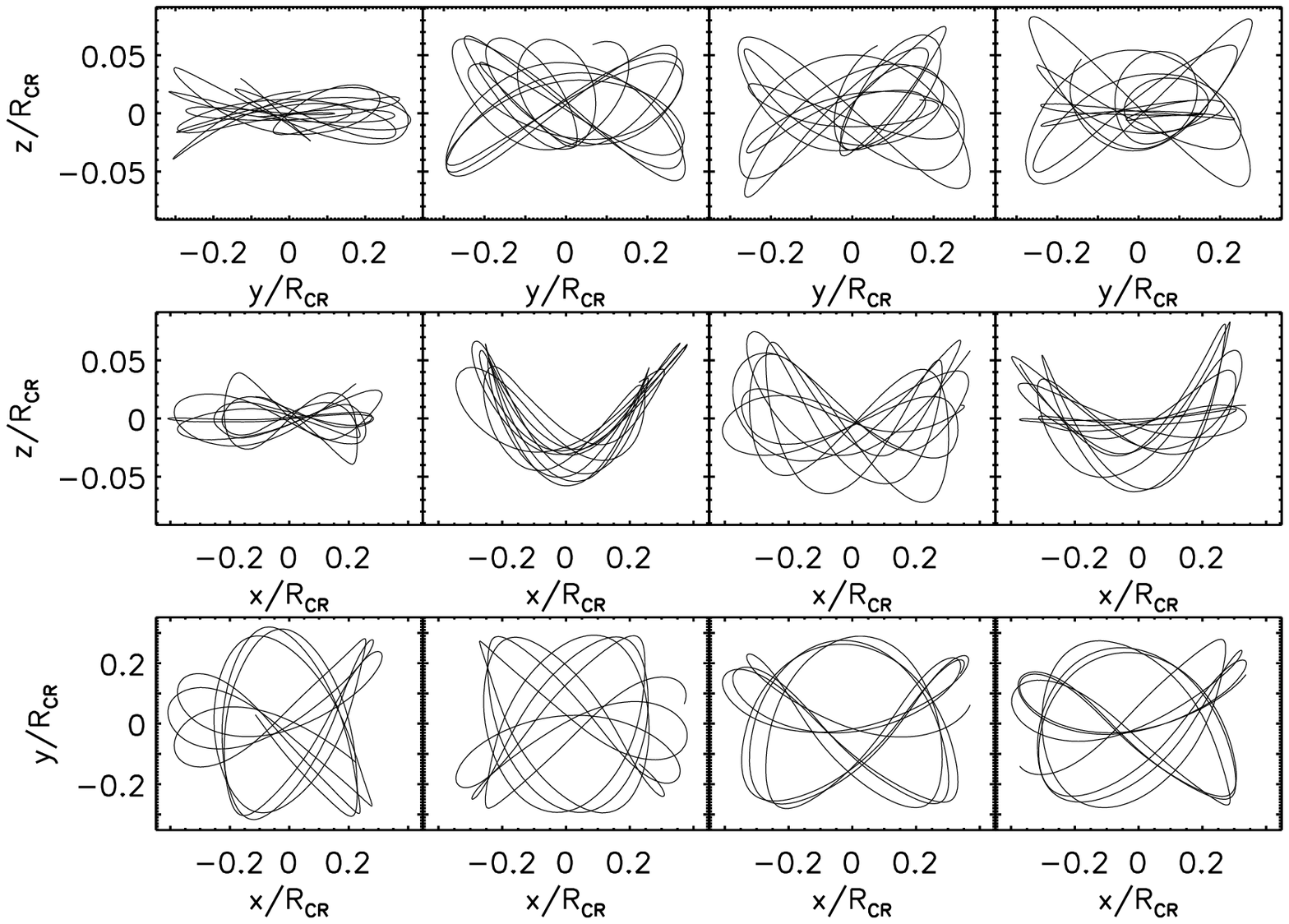}
\includegraphics[height=0.35\textwidth]{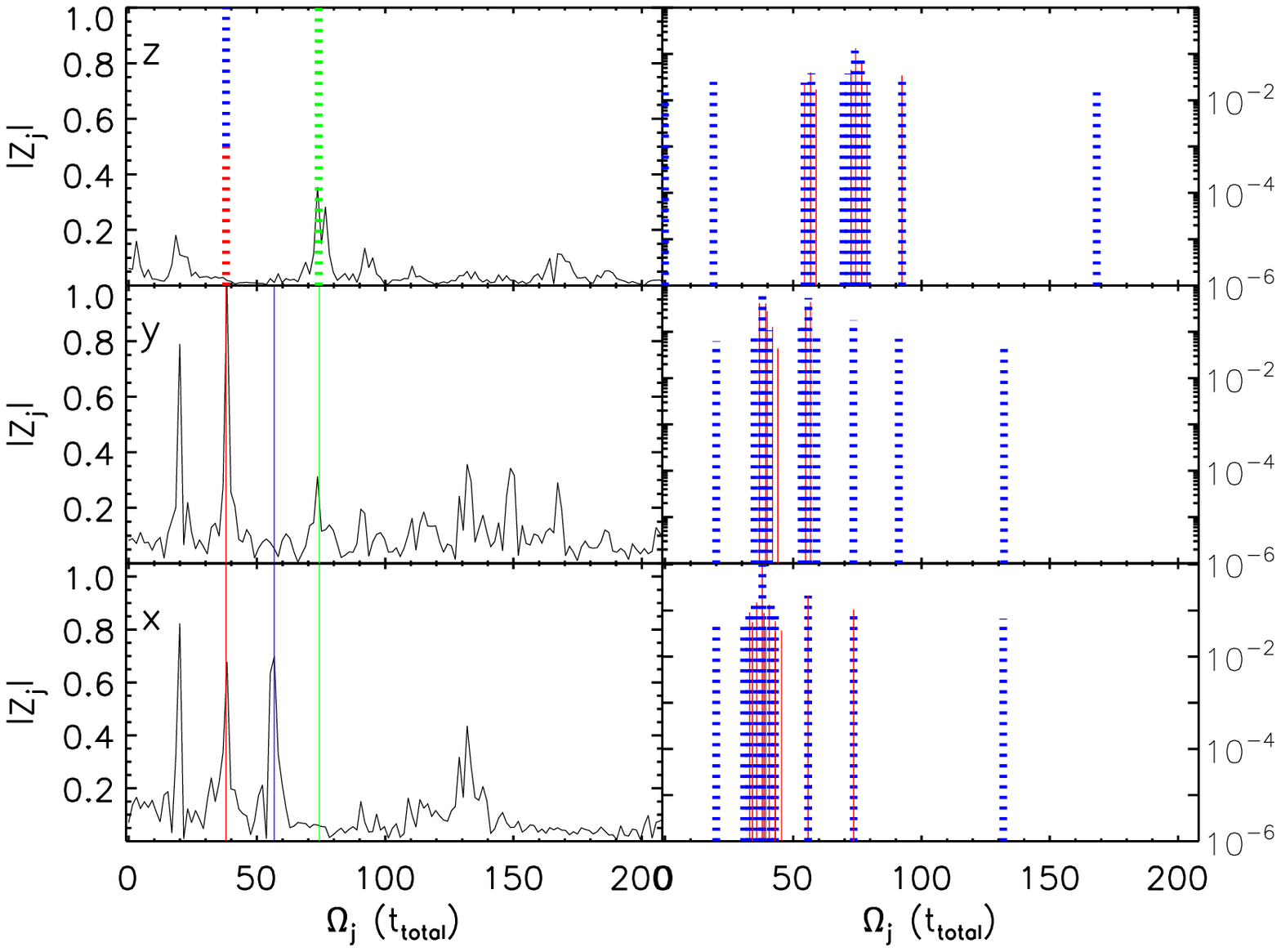}
\caption{ $Left$ set of panels: orbit 2745. From top to bottom, the
  orbit in y-z, x-z and x-y planes, respectively. From left to right,
  the result for time 6.005-7.024, 7.0245-8.048, 8.0485-9.072,
  9.0725-10.096Gyr. Note that the scales of the ordinates and abscissas
  are not always the same, so as to allow for a better
  resolution. $Right$ set of panels: FFT spectrum for this orbit 
  (left) and the extracted lines (right). From top to bottom, the
  results for the z, y and x components, respectively. First column of
  the right set of panels: The red, blue and green solid lines denote
  the positions of $\omega_1$, $\omega_2$ and $\omega_3$ from NAFF,
  while the corresponding dashed lines represents the main frequencies
  from CA method. The amplitude of the spectrum is normalized by the
  largest amplitude among three components. Second column of right set of
  panels panel: The red solid and blue dotted lines represent the
  extracted lines from NAFF and CA, respectively.
  $[\omega_1,\omega_2,\omega_3]$ =[37.980, 56.760, 74.362 ] (NAFF) and
  $[\omega_1,\omega_2,\omega_3]$ =[37.997, 38.004, 74.104 ] (CA) . The
  ordinate of the rightmost panel is  in logarithmic scale.}\label{fig:orbit2745}
\end{center}
\end{figure*}

The rightmost panels of Figure ~\ref{fig:orbit2745} show 10 lines extracted by
 the CA  method with FMFT and 10 extracted by the NAFF method in the spectra of the three $(x, y, z)$  
 components for orbit 2745. It is seen that most lines from the two methods agree, but some lines are significantly different. It is also noted that the primary frequency in the CA code is found by the largest amplitude $|Z_{j,pv}|$ (defined below eq. 2) in the FFT spectrum, which is slightly different from that done in the NAFF method. For most orbits, the frequency with the largest amplitude $|Z_{j,pv}|$ is consistent with the frequency with the largest amplitude among $|Z_{j,p}|$ and $|Z_{j,v}|$. However, for some orbits this is not true.  In Table~\ref{table:1315}, we show the frequencies and amplitudes of the first 20 strongest lines in the FFT spectra of orbit 1315.  It is seen that the frequency with the largest  $|Z_{j,pv}|$ is 78.242571, while the frequency with the largest amplitude among $|Z_{j,p}|$ and $|Z_{j,v}|$ is 119.665109 in the x component. 
 


We refer the interested readers to CA98 for a full description
of their technique.  Here we only give a brief overview and some
modifications on the new version of their code.  There is a clean
distinction between the main and fundamental frequencies in this new
version.  The main frequencies are the frequencies whose amplitudes
are the maximum  (or second maximum) on each coordinate. These
frequencies are used to determine whether or not the orbit
is resonant. The fundamental frequencies are  the independent frequencies. We will take an example to illustrate this difference.  If there is no integer non-zero vector $(l,m,n)$ to satisfy     
 $l\omega_1+m\omega_2+n\omega_3=0$, these main frequencies are
independent. If the rest of the spectral lines can be expressed as the
linear combinations of them, then the fundamental frequencies are the same
as the main frequencies. If there are more than three independent
frequencies, the number of fundamental frequencies will be 4, and thus
the orbit is classified as an irregular type in CA. If there is one
resonance, then the three main frequencies are not independent, the
main frequencies are not the fundamental frequencies.

\section{Main frequencies in NAFF and CA}
There are two different  conceptual frequencies in the literature, one is the fundamental frequency, the other is the main frequency. Unfortunately, these two
are sometimes confusingly used. 

In NAFF,  the fundamental frequencies are frequencies of the angle variables in the case of a regular orbit for which the action/angle variables exist.
In that case any coordinate time series will have a
spectrum made of discrete lines at frequencies that can be written as
linear combinations with integer coefficients of three 
independent ``fundamentals frequencies". However, unless the coordinates
used are close to angle variables, there is no reason
why the dominant line in one spectrum should be one of those fundamental
frequencies. For box orbits, the fundamental frequencies are identified by the highest amplitude terms  in the Cartesian coordinates. On the other hand, for tube orbits, 
the terms with the second or subsequent  highest amplitudes  are taken as the fundamental frequencies  ~\citep{1998ApJ...506..686V}.

In the CA method, the main frequencies are frequencies with the maximum or subsequent highest amplitudes of each coordinate, which is the same as the ``fundamental" frequencies in NAFF.
The main frequencies in CA are used further to determine whether or not there are resonances. If there is no resonance among the main frequencies, they may be taken as fundamental frequencies too.
If, however, there are resonances, then the main frequencies are used to determine one to three linearly  independent fundamental frequencies for regular orbits, or more than three for irregular orbits. 
 Therefore, the main frequencies in CA coincide with the ``fundamental" frequencies in NAFF.
In the remaining of the paper, we will use the ``main" frequencies and ``fundamental" frequencies as defined in CA.

The first step to get the main frequencies is to extract the lines from
the Fourier spectra. We use both the position and velocity components
$X(t)+iV(t)$ to get the spectrum for each component. In order for
positions and velocities to contribute in a comparable manner, we use
a normalised, dimensionless position and velocity to do the Fourier
transform. The original position and velocity are divided by $R_m$ and
$V_m$, respectively, where $R_{m}=\sqrt{\langle x^2+y^2+z^2\rangle}$
and $V_{m}= \sqrt{\langle v_x^2+v_y^2+v_z^2\rangle}$, $x$, $y$ and
$z$ are the three positions, $v_x$, $v_y$ and $v_z$ are the three
velocities and $\langle \cdot\cdot\cdot \rangle$ denotes the average
over time along the orbit.

The detailed method to  extract the spectrum is given in ~\cite{2003math......5364L}, we refer the reader to his paper for further details. 
Here we just point out that the strongest spectral lines in each component are obtained using an accurate numerical technique. 
In NAFF, all extracted lines are sorted by amplitude in descending
order. The first main frequency $\omega_1$ corresponds to the line
with the largest amplitude, the second main frequency $\omega_2$ is
the next highest peak coming from a different component and a value
different from the first main frequency. 
The third main frequency is
 one of the remaining frequencies, should come from the remaining
 component, and should not be any linear  
 combination of $\omega_1$ and $\omega_2$.

In CA, the $x$, $y$ and $z$ axes should be aligned
with the major, intermediate and minor axes
of the system. Then the main frequency from each component should
yield $\omega_x<\omega_y<\omega_z$, where $\omega_x$, $\omega_y$ and
$\omega_z$ are the highest peaks from the spectrum of the $x$, $y$ and
$z$ components, respectively. Therefore, if the frequency from the
largest peak in each component does not satisfy
$\omega_x<\omega_y<\omega_z$, the CA method switches the
corresponding coordinates, unless the two corresponding amplitudes are
very close to each other.
The first main frequency is the smallest frequency among $\omega_x$,
$\omega_y$ and $\omega_z$. The second and third main frequencies are
from the frequency components with intermediate and largest values
among $\omega_x$, $\omega_y$ and $\omega_z$, respectively. In principle,
when $\omega_x<\omega_y<\omega_z$, then the second main
frequency is from the spectrum of the $y$ component, and the third
main frequency is from the spectrum of the $z$ component. 
However, in practice,  when $\omega_y$ is quite close to $\omega_x$, then $\omega_2$ is searched in descending order of
amplitude in the $y$ spectrum until $\omega_2$ is significantly larger than
$\omega_1$. A similar treatment is adopted for the third main frequency.  

\begin{figure}
\centering
\includegraphics[angle=0, width=80mm]{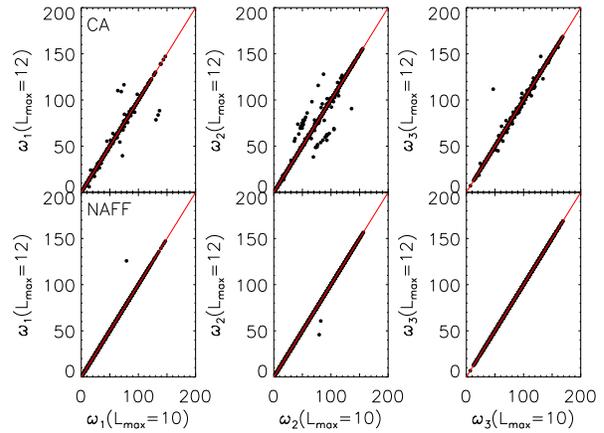}
\caption{Comparison of main frequencies from different values of $L_{\rm max}$. The solid line represents equality
of two frequencies for two different parameters  $L_{\rm max}$ (12 and 10).  The top and bottom panels represent the results from the CA and NAFF methods, respectively. The sample interval is $t_{\rm total} $.} 
\label{fig:mf_comp_Lmax}
\end{figure}
\begin{figure}
\includegraphics[angle=0, width=80mm]{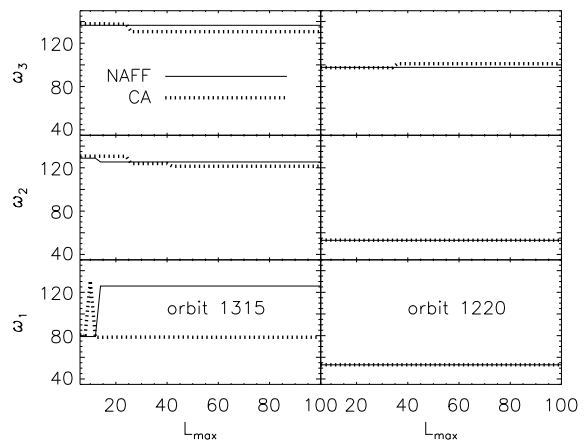}
\caption{Dependence of  three main frequencies on $L_{\rm max}$ for both NAFF (solid line) and CA (dotted line) for orbit 1315 (left panel) and orbit 1220 (right panel). 
The top, middle and bottom panels represents the results for $\omega_3$, $\omega_2$ and $\omega_1$, respectively. For orbit 1220, $\omega_1$ and $\omega_2$ from two methods are same.} 
\label{fig:Lmax}
\end{figure}

Since the main frequencies are selected among the extracted lines in
the spectrum in both methods, they may depend on the candidate number
of the extracted lines $L_{\rm max}$.  Figure~\ref{fig:mf_comp_Lmax}
shows a comparison of the main frequencies obtained with $L_{\rm
  max}=10$ and with $L_{\rm max}=12$ for both methods. It is seen that
only a small number ($<0.1\%$) of the main frequencies in NAFF have
been changed when using different values of $L_{\rm max}$, while about
$6\%$ of the main frequencies have been changed in CA.  Here the
changes in the main frequencies from $L_{\rm max}=10$ to $L_{\rm
  max}=12$ mean that  the largest frequency difference of  $|\omega_i
(L_{\rm max}=12)/\omega_i (L_{\rm max}=10)-1|$ $(i=1,2,3)$ is larger
than 0.01. It is easy to understand these changes of the main
frequencies with the increasing number of  $L_{\rm max}$ in both the
NAFF and CA methods. In the CA method, the frequencies and amplitudes
of the extracted lines are corrected by the next extracted lines,
therefore, the frequency and amplitude from the extracted lines are
changed when $L_{\rm max}$ is different. In the NAFF method, the
increasing number of $L_{\rm max}$ may give new frequencies and
amplitudes. To illustrate this, we show the frequencies and amplitudes
of the extracted lines in orbit 1315 for both the NAFF and the CA
methods in Table~\ref{table:1315peak}. From the definition of the main
frequencies in the two methods, we know that $\omega_1=79.181$ ($k=3$ in the x component) with $L_{\rm max}=10$, 
and  $\omega_1=125.802$ with $L_{\rm max}=12$ $(k=11)$ in NAFF. The shift  of $\omega_1$ in NAFF is because
a new line with a large amplitude is found in the eleventh step. In the CA code, the increasing number of $L_{\rm max}$ changes the frequencies and amplitudes, 
therefore, $\omega_1$ can be changed. In Figure~\ref{fig:Lmax}, we show the dependence of three main frequencies on the value of $L_{\rm max}$ for orbit 1315 (Left) and orbit 1220 (Right) .
It is seen that the main frequencies from NAFF will not be changed if   $L_{\rm max}\ge 12$, 
while there is a small fluctuation along the $L_{\rm max}$ value for orbit 1315. For orbit 1220, only $\omega_3$
has been changed at $L_{\rm max}=36$, and will be kept as a constant with $L_{\rm max}>36$. 
In order to avoid the missing lines and save the compute time, we adopt $L_{\rm max}=12$ in the remainder of the paper unless stated otherwise.     

\begin{figure}
\includegraphics[angle=0, width=80mm]{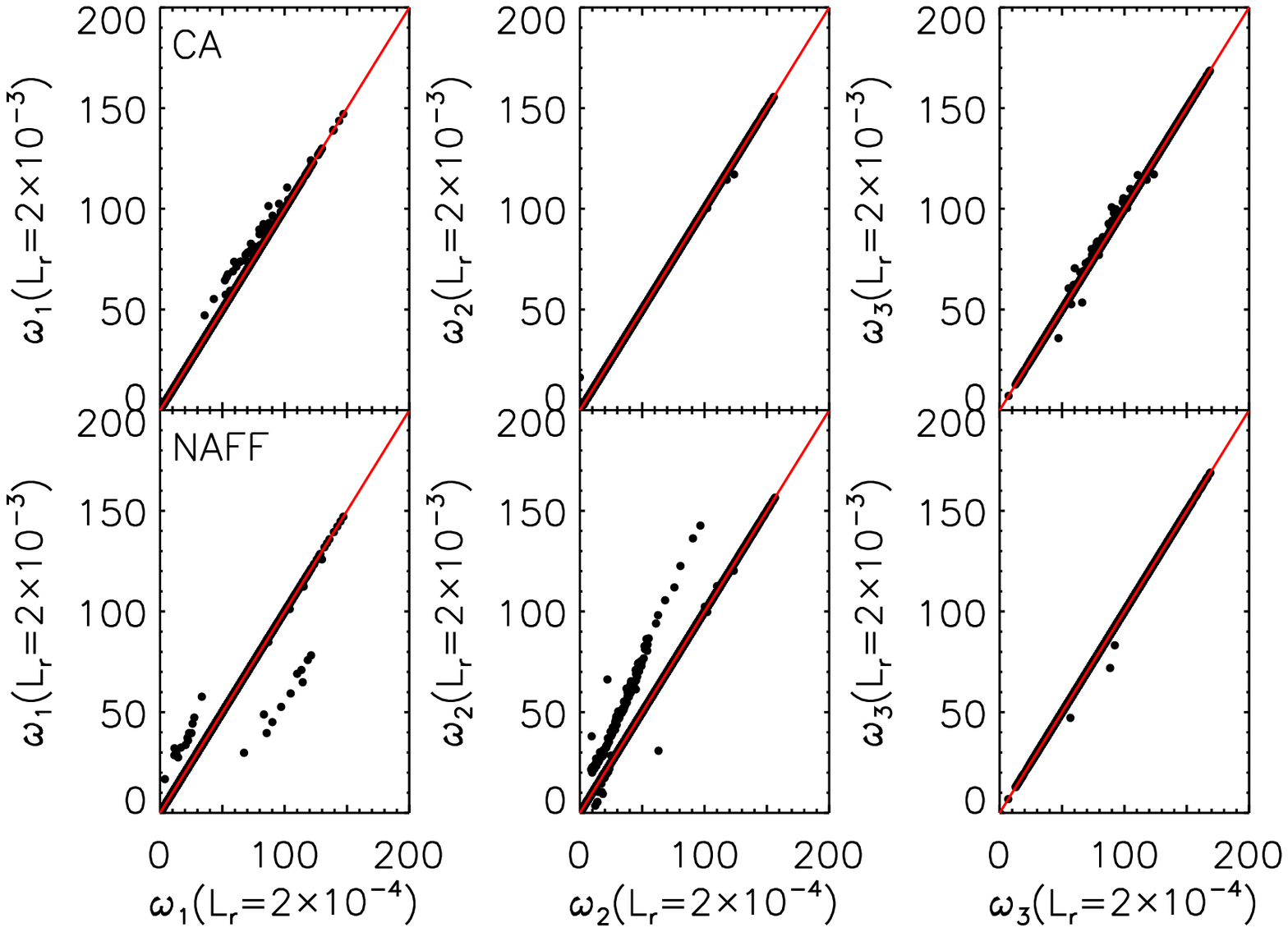}
\caption{Comparison of main frequencies from different values of
  $L_{\rm r}$. The red solid line represents equality of  two
  frequencies for two different parameters $L_{\rm r}$.  The top and
  bottom panels represent the results from the CA and NAFF methods,
  respectively. The time interval used here is  $t_{\rm total} $.}
\label{fig:mf_comp_Leps}
\end{figure}

In NAFF, the absolute difference between the first and second
main frequencies $|\omega_i-\omega_j|  (i=1,2,3, j=1,2,3$,  and $i\ne j$)
must be  larger than a critical value  $L_{\rm r,a}$, which we define as the critical absolute frequency difference.  In CA, the
parameter to distinguish two frequencies  is similar to NAFF, but with
the value of the relative frequency difference  $|\omega_i-\omega_j|/\sqrt{{\omega_i}^2+{\omega_j}^2}$
larger than a critical value $L_{\rm r}$.  In order to compare the
main frequencies in these two methods, we introduced a definition consistent
with that of CA,
i.e. $|\omega_i-\omega_j|/\sqrt{{\omega_i}^2+{\omega_j}^2}>L_{\rm
  r}$.  In Figure~\ref{fig:mf_comp_Leps}, we show the comparison of
main frequencies from $L_{\rm r}=2\times 10^{-4}$ and  $L_{\rm
  r}=2\times 10^{-3}$, where the first one is suggested by the CA
method. We found that around $6\%$ and $1.5\%$ of the orbits have a
different main frequency in the NAFF and the CA methods,
respectively. Here we define two main frequencies as different if
$|\omega_i (L_{\rm r}=2\times 10^{-4})/\omega_i ( L_{\rm
  r}=2\times 10^{-3})-1|$ $(i=1,2,3)$ is larger than 0.01.
  
\begin{figure}
\includegraphics[angle=0, width=80mm]{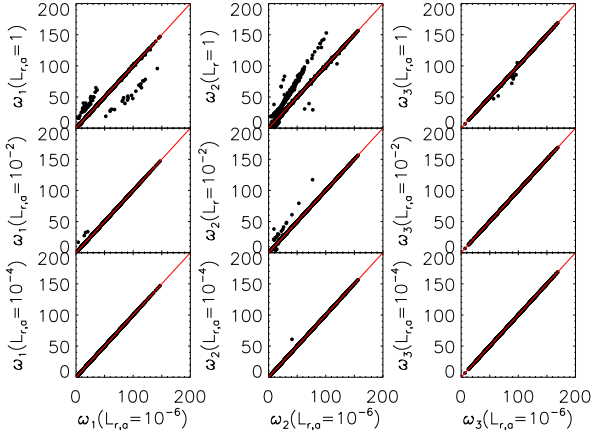}
\caption{Comparison of main frequencies from different values of $L_{\rm r,a}$ in NAFF.  The solid line represents equality of two frequencies for two parameters  $L_{\rm r,a}$.   The sample interval is $t_{\rm total} $.}
\label{fig:mf_comp_Leps_abs}
\end{figure}

\begin{figure}
\includegraphics[angle=0, width=80mm]{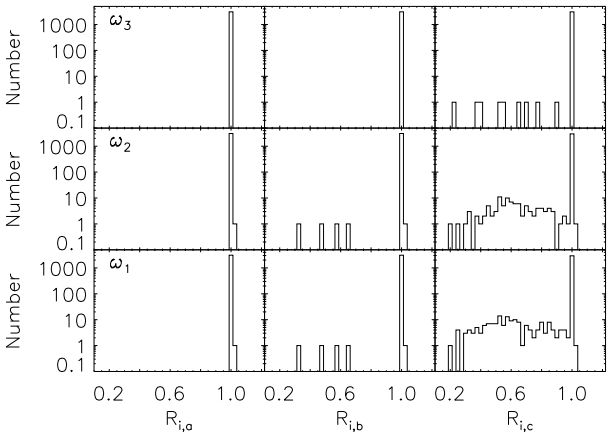}
\caption{The distribution of the amplitude ratios. From top to bottom,
  the results are for $\omega_3$, $\omega_2$ and $\omega_1$,
  respectively. The sample interval is $t_{\rm total} $. The
  parameters $R_{i,a}$, $R_{i,b}$ and $R_{i,c}$ are defined in
  eqs.(~\ref{eq:ra})-(\ref{eq:rc}). Note that the scale for the ordinate is logarithmic.}
\label{fig:ratio_amp_naff_his}
\end{figure}

Since our orbits are extracted from a simulation, they are necessarily
much noisier than those obtained from an analytic potential. In order
to estimate this effect on the main frequency detection, we will
vary the absolute critical value $L_{\rm r,a}$ in the NAFF
method, to check whether any lines with very small amplitude are taken as the main frequencies. 
In Figure ~\ref{fig:mf_comp_Leps_abs}, we compare the main frequencies with different $L_{\rm r,a}$,
and find that even when the value of $L_{\rm r,a}$ is increased from $10^{-6}$ to 1, only $22.3\%$ of
the orbits change their main frequencies:  the largest frequency change of $|\omega_i (L_{\rm r,a}=1)/\omega_i ( L_{\rm r,a}=2\times 10^{-6})-1|>0.01$.
We define nine parameters to describe the corresponding amplitude variation: 

\begin{equation}\label{eq:ra} 
R_{i,a}=A_i( L_{\rm r,a}=10^{-4})/A_i( L_{\rm r,a}=10^{-6}) , 
\end{equation}
\begin{equation}\label{eq:rb}
R_{i,b}=A_i( L_{\rm r,a}=10^{-2})/A_i( L_{\rm r,a}=10^{-6}) , 
\end{equation}
\begin{equation}\label{eq:rc}
R_{i,c}=A_i( L_{\rm r,a}=1)/A_i( L_{\rm r,a}=10^{-6}) , 
\end{equation}
where $i=1,2,3$ and $A_i$ are the amplitudes of the main frequencies
$\omega_i$. In Figure ~\ref{fig:ratio_amp_naff_his}, we show the
distribution of these parameters $R_{i,a}$,  $R_{i,b}$ and $R_{i,c}$ in the NAFF method.
We can see that some lines with low amplitude appear as new main
frequencies as $L_{\rm r,a}$ increases. However, the number of orbits for which the ratio of the amplitudes is considerably different from unity is quite small. This is true even when we change
this parameter by 6 orders of magnitude, from $10^{-6}$ to 1 
(rightmost panels). In other words, the simulation noise does not affect the
  main frequencies significantly.
  
We also check the distribution of the amplitude ratios in the CA method, and find results similar to those in NAFF.
Therefore, the effect of the critical parameter to distinguish two
frequencies is small in both methods.  However, in the CA method, if we increase the value of  $L_{\rm r}$,
the number of fundamental frequencies may be changed significantly,
which increases the fraction of regular orbits significantly. 
Indeed in CA, the parameter $L_{\rm r}$ has two meanings: One is the frequency difference, which is the same as shown in our paper,
while the other is the critical value determining whether an orbit is resonant, or not. In the CA code,
if $|l\omega_1+m\omega_2+n\omega_3|/\sqrt{(l\omega_1)^2+(m\omega_2)^2+(n\omega_3)^2} $ is smaller than 
$L_{\rm r}$, then a resonance has been found. Since the number of the fundamental frequencies depends on the
resonance number of the orbits, $L_{\rm r}$ can affect the number of fundamental frequencies.
In order to give a more detailed comparison of the main frequencies between the NAFF and CA methods, 
we adopt a relative critical value $L_{\rm r}=2\times 10^{-4}$ in the remainder of the paper, unless otherwise indicated.



\begin{figure}
\includegraphics[angle=0, width=80mm]{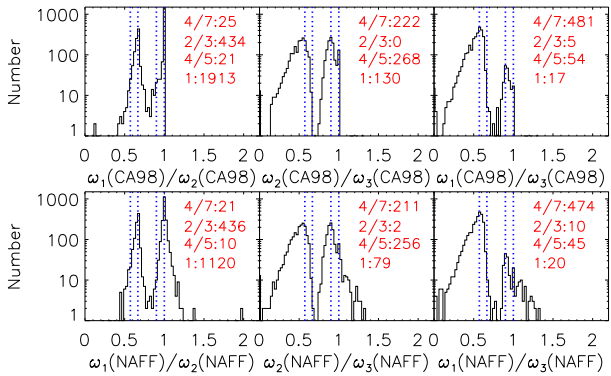}
\caption{Histogram of the ratios of the main frequencies from the CA (top) and NAFF  (bottom) methods. From left to right,  
the results are for the ratios with $\omega_1/\omega_2$,  $\omega_2/\omega_3$, and $\omega_1/\omega_3$, respectively.
From left to right, the vertical dashed lines represent ratios with 4/7, 2/3, 4/5 and 1, respectively. The sample interval is $t_{\rm total} $. 
The ratio and the corresponding orbit numbers are indicated in the top-right corner of each panel.}
\label{fig:mf_dis}
\end{figure}

Figure~\ref{fig:mf_dis} shows the histogram of the ratios of the three
main frequencies from both the NAFF (bottom) and CA (top) methods. It
is seen that there are typical peaks in these distributions, which
indicate the intrinsic orbit types in our N-body bar. Note that the
ordinate is in a logarithmic scale, which means that the peaks are
very high, i.e. that many orbits are in families with well defined
frequency ratios. For both methods, the face-on view, ($x$, $y$), has
two clear peaks. The highest peak is
for 1:1, and the second highest 
for 2:3, the two having amplitudes of 1120 and 436 in NAFF, 1913 and 434 in CA98.  The two
edge-on views, ($y$, $z$) and ($x$, $z$), also have two clear
peaks one at 4:5 and the other at
4:7. The implications of this result will be discussed elsewhere.

\begin{figure}
\includegraphics[angle=0, width=80mm]{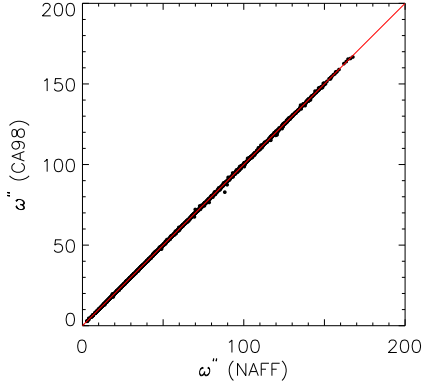}
\caption{Comparison of one similar main frequency  from  NAFF  with that from CA. The solid line represents equality of two frequencies from these two methods. The sample interval is $t_{\rm total} $.}
\label{fig:mf_comp_1}
\end{figure}

We define 9 parameters to check whether a main frequency agrees in the two methods:   

\begin{equation}
\delta f_{1}^{ij}=\bigg| \frac{\omega_i \rm(NAFF)}{\omega_j \rm(CA)}-1\bigg|,
\end{equation}
with $i=1,2,3$ and $j=1,2,3$. We take the minimum value of $\delta f_{1}^{ij}$ as  $\delta f_1$. If $\delta f_1$ is small, then at least one main frequency from CA is consistent with one from NAFF.  
The corresponding  frequency to $\delta f_1$ is defined as $\omega^{\prime\prime}$.   
We find that $97\%$ of the orbits have the minimum  $\delta f_{1}$ smaller than 0.01, in other words, 
these orbits have at least one similar main frequency from the two methods, which can also be seen from the comparison of one similar main frequency in
Figure~\ref{fig:mf_comp_1}. Then we define 18 parameters to check two main frequencies agree from two methods, these parameters are given as
\begin{equation}
\delta f_{2}^{ij,{i^\prime}{j^\prime}}=\bigg| \frac{\omega_i\rm(NAFF)}{\omega_j \rm(CA)}-1\bigg|+\bigg| \frac{\omega_{i^\prime }\rm(NAFF)}{\omega_{j^\prime} \rm(CA)}-1\bigg|,
\label{eq:2agree}
\end{equation}
with $i=1,2,3$, $j=1,2,3$, $i^\prime=1,2,3$, $j^\prime=1,2,3$, and $i^\prime\neq i$, $j^\prime\neq j$.    
The minimum value of $\delta f_{2}^{ij,{i^\prime}{j^\prime}}$ is defined as $\delta f_{2}$. Two main frequencies agree in the two methods if  $\delta f_{2}$ is smaller than
a critical value $\delta f_{2,0}$.  If the remaining main frequencies from two methods 
are also close (the frequency difference is smaller than $\delta
f_{2,0}/2$), then all three main frequencies are in agreement.  
The first two identical main frequencies are defined as $\omega_1^{\prime\prime}$ and 
$\omega_2^{\prime\prime}$, and the remaining main frequency as $\omega_3^{\prime\prime}$. We find $88\%$ orbits have two frequencies in agreement and $39\%$ orbits have 
three frequencies in agreement from two methods if  $\delta f_{2,0}=0.02$. If we increase $\delta f_{2,0}$ to 0.1, then $99\%$ of the orbits have two frequencies in agreement and $80\%$ of the orbits have 
three frequencies in agreement between the two methods. In other
words, most orbits have an average difference in the main frequency
smaller than $5\%$.

\begin{figure}
\includegraphics[angle=0, width=80mm]{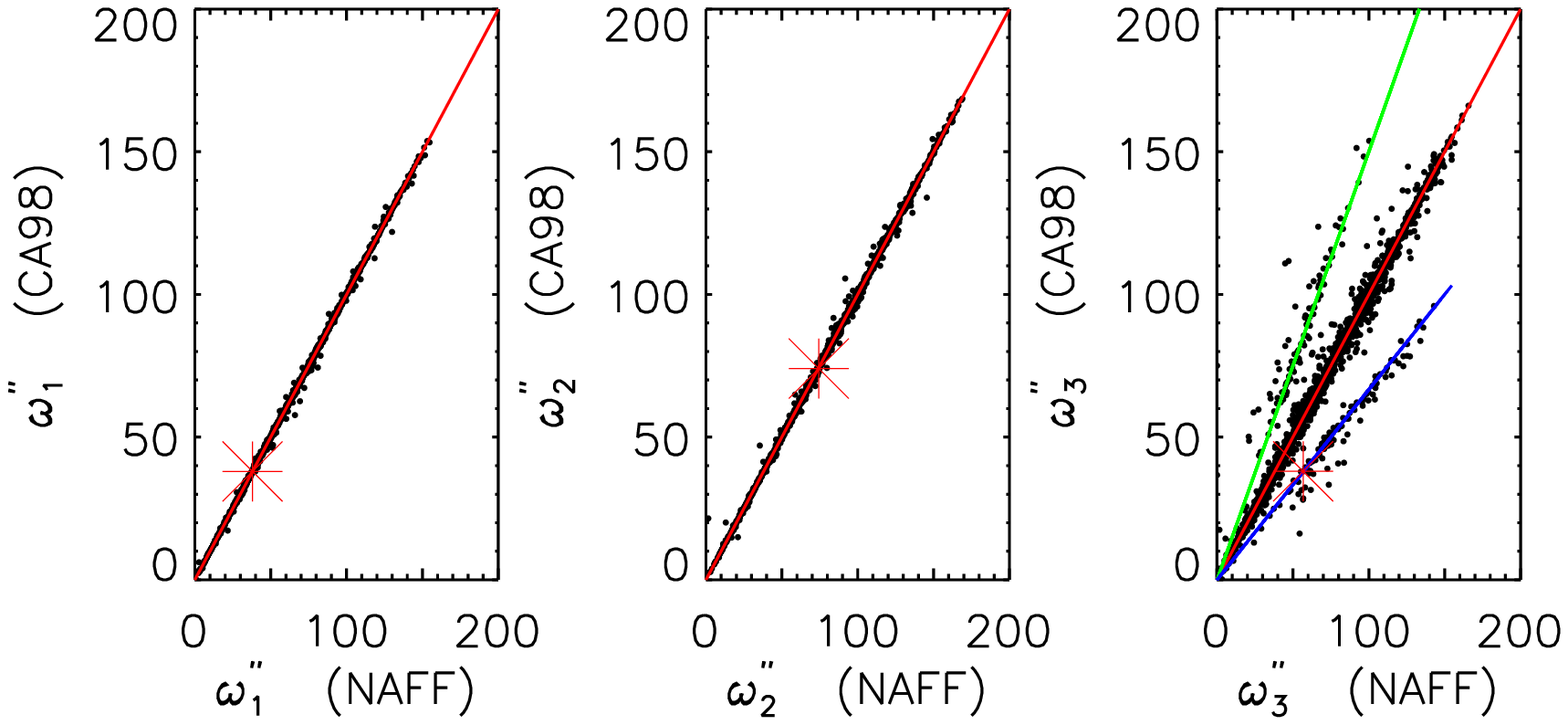}
\caption{Comparison of the three main frequencies from the NAFF and CA methods. The red solid line represents the equality of two frequencies from the two methods.
From left to right, the results are for $\omega_1^{\prime\prime}$, $\omega_2^{\prime\prime}$ and $\omega_3^{\prime\prime}$, respectively.   
In the right panel, the green and blue solid lines present $Y=AX$ with $A=3/2$ and $A=2/3$, respectively. The sample interval is $t_{\rm total} $. The red asterisk symbol denotes the location of  orbit 2745 in each panel. }
\label{fig:mf_comp_3}
\end{figure}

In Figure~\ref{fig:mf_comp_3}, we show the comparison of the three
main frequencies from the two methods. The two first panels,
  referring to the two first frequencies, show a close equality,
  with all points distributed very close to the diagonal. The third
  panel, referring to the third frequency, has a different
  structure. About $80\%$ of the points (2458 orbits) are around the diagonal, but with a considerably larger
  spread than for the first and second frequency. This may argue that
  this third frequency is less accurately defined than the other
  two. Note also that a considerable number of orbits  (144 orbits, $5\%$) are located at the wings along the green and blue
  solid lines which follow $Y=AX$ with $A=3/2$ (green) or $A=2/3$
  (blue), respectively. This could be due to a badly 
recognised third frequency.

 \begin{table*}
\caption{Frequencies and amplitudes extracted from the spectrum of orbit 1315 in the x (top), y (middle) and z (bottom) components using the NAFF and CA methods with different $L_{\rm max}$. }\label{table:1315peak}
\begin{center}
\begin{tabular}{cclllllll}\hline
 k& $\omega^\prime_k $ (NAFF) & $A^\prime_k$ (NAFF) & $\omega^\prime_k $ (NAFF) & $A^\prime_k$ (NAFF) & $\omega^\prime_k $ (CA) & $A^\prime_k$ (CA) & $\omega^\prime_k $ (CA) & $A^\prime_k$ (CA)\\  
   &$L_{\rm max}=10$ & $L_{\rm max}=10$ &  $L_{\rm max}=12$  & $L_{\rm max}=12$ & $L_{\rm max}=10$  & $L_{\rm max}=10$ &  $L_{\rm max}=12$  & $L_{\rm max}=12$ \\    
  \hline\hline

1   &118.646975 & 0.087493 & 118.646975         & 0.087493     &     123.906921   &    0.205454 &  123.903713    &  0.204517     \\
2   &118.646975  & 0.129730       & 118.646975  & 0.129730     &       80.383856    &   0.197367  &    80.379904    &  0.197389     \\
3   &79.180676    & 0.189038       &  79.180676    & 0.189038     &     130.690324   &    0.222754 &  130.526476    &  0.204632                              \\
4   &130.964723  & 0.178945       &  130.964723  & 0.178945    &     127.104809    &   0.190681 &   127.053323   &  0.188998     \\
5   &79.180569    & 0.187178        &  79.180569    & 0.187178    &       78.497611    &   0.215303 &     78.493748   &  0.215055     \\
6   &83.774549    & 0.120298        &  83.774549    & 0.120298    &     121.152373    &   0.180770 &   121.152083   &  0.180511       \\
7   &76.471528    & 0.057949        &  76.471528    & 0.057949    &     132.023087    &   0.174899 &    132.228818  &  0.193995   \\
8   &122.210680  & 0.178263        &  122.210680  & 0.178263    &      83.047291    &   0.134492  &      83.037662  &  0.134594    \\
9   &131.004247  & 0.159953        &  131.004247  & 0.159953    &     119.680380   &   0.142824  &   119.677448  &   0.142573       \\
10 &134.123148  & 0.079459       &  134.123148   &0.079459    &       77.055611    &  0.114834     &      77.051043  &  0.114728       \\
11 &                        &                          &  125.801781  &0.195909    &                                &                     &    133.928603  &  0.104530  \\
12 &                        &                          &  124.280162  &0.097235    &                                &                     &      87.622990  &  0.072769 \\
\hline\hline
1  &  120.187879   &  0.289342     &  120.187879   &  0.289342   &124.332263   &  0.447271   & 124.359786   &  0.443748 \\
 2  & 132.703213   &  0.276719     & 132.703213   &  0.276719    &130.661153    &  0.476913   &   130.663652  &   0.476665 \\
 3  & 134.960731   &  0.192684     & 134.960731   &  0.192684   & 127.359892   &  0.430963    &     127.374838  &   0.430598 \\
 4  &  134.654657  &  0.222301     &  134.654657  &  0.222301   &  121.552062  &    0.403108   &    121.646264  &   0.390782 \\ 
 5  &  128.870117  &  0.408115     &  128.870117  &  0.408115   &   132.318424   &  0.453290    &   132.320529   &  0.453740 \\ 
  6 &  131.047180  &  0.334746     &  131.047180  &  0.334746   &     119.660980  &   0.439960   &    119.716928   &  0.440926\\  
  7 &  125.360971  &  0.463758     &  125.360971  &  0.463758   &   133.982798   &  0.242381    &       133.982529  &   0.242588 \\
  8 &  127.308115  &  0.133632     &  127.308115  &  0.133632   &   118.283853   &  0.248884    &     118.023337  &   0.341200 \\
  9 &  122.980739  &  0.414632    &  122.980739  &  0.414632    &    80.485974    &  0.132293    &   80.146343  &   0.128368\\   
10 &  119.665109  &  0.304908    &  119.665109  &  0.304908   &   78.780880    &    0.109954    &  78.661243   &  0.110079\\
11 &                         &                        &  121.532408   & 0.054014   &                           &                        &    116.738274  &   0.126798\\
12 &                         &                        &   124.217988   & 0.081351  &                           &                       &       83.026349   &  0.084185\\
\hline\hline
 1  & 129.788437        &   0.160866   & 129.788437        &   0.160866 & 138.000765   &  0.345491 &  138.002662 &    0.343542\\   
 2  & 127.865342        &   0.097260   & 127.865342        &   0.097260  &   135.090665  &   0.291035 &  135.135194   &  0.290528\\
 3  & 128.132109        &   0.182508   & 128.132109        &   0.182508   &    140.857289  &   0.264038  & 140.859943  &   0.263717\\
 4  & 132.335399        &   0.179940    & 132.335399        &   0.179940  &    143.924934   &  0.210984 &  143.926056   &  0.210835\\
  5  & 131.938453        &   0.189985   & 131.938453        &   0.189985  &    130.765206  &   0.327860 &  130.706509 &    0.327987\\ 
  6  & 136.540957        &   0.342655   & 136.540957        &   0.342655  &      132.723722  &   0.244343  & 132.761358   &  0.232638\\
  7  & 133.705729        &   0.138481     & 133.705729        &   0.138481 &      146.306622  &   0.202305  & 146.308937  &   0.202298\\
  8  &  139.613950        &   0.325237     &  139.613950        &   0.325237 &  129.506709  &   0.184226  & 129.170063  &   0.259542\\   
  9  &  137.906023        &   0.064947      &  137.906023        &   0.064947 & 147.877606  &   0.156429 &  147.878724  &    0.156423\\   
 10 &  146.398331        &   0.155171     &  146.398331        &   0.155171 &   110.364005 &    0.101131&   110.355605 &     0.101467\\  
  11 &                               &                         &   146.398154       &   0.192978 &                          &                       &  116.926172 &    0.094574\\
  12  &                              &                         &    142.327606      &   0.213880 &                         &                        &     127.811826 &    0.115941\\

 \hline  
 
   \hline
 \end{tabular}
\end{center}

\end{table*}

 \begin{table*}
\caption{Frequencies and amplitudes of the first 20 strongest lines in the FFT spectra of orbit 1315. From top to bottom,  the results for x , y  and z  components, respectively. 
The amplitude is normalized by the largest values of $|Z_{j,pv}|$, $|Z_{j,p}|$ and $|Z_{j,v}|$ in the x, y and z components.}\label{table:1315}
\begin{center}
\begin{tabular}{cclllllll}\hline
 & $\Omega_j $ & $|Z_{j,pv}|$ & $\Omega_j $  & $|Z_{j,p}|$   & $\Omega_j $  & $|Z_{j,v}|$  \\
 \hline\hline
  1&   78.242571 &    0.376261 &   78.242571 &    0.397491 &  119.665109 &    0.441276 \\
     2&   79.776739 &    0.341673 &  131.938454 &    0.388476 &   79.776739 &    0.372881 \\
     3&  131.938454 &    0.341376 &   76.708403 &    0.341961 &   78.242571 &    0.353760 \\
     4&  119.665109 &    0.322304 &  133.472622 &    0.330353 &  124.267613 &    0.323890 \\
     5&  118.130941 &    0.306547 &  121.199277 &    0.310250 &  127.335949 &    0.322616 \\
     6&  130.404285 &    0.290761 &   79.776739 &    0.307313 &  118.130941 &    0.321816 \\
     7&  133.472622 &    0.277273 &  118.130941 &    0.290476 &  130.404285 &    0.309150 \\
     8&   76.708403 &    0.261478 &  116.596773 &    0.283636 &   82.845075 &    0.303582 \\
     9&  124.267613 &    0.260451 &  130.404285 &    0.271128 &  131.938454 &    0.286639 \\
    10&  127.335949 &    0.230389 &  122.733445 &    0.241251 &   81.310907 &    0.239860 \\
    11&  121.199277 &    0.222195 &  115.062605 &    0.190504 &  125.801781 &    0.220963 \\
    12&   82.845075 &    0.216798 &  124.267613 &    0.175397 &  133.472622 &    0.211252 \\
    13&  122.733445 &    0.206655 &   87.447580 &    0.144521 &  135.006790 &    0.203641 \\
    14&  116.596773 &    0.203209 &  119.665109 &    0.114172 &   75.174235 &    0.171446 \\
    15&   81.310907 &    0.171751 &   75.174235 &    0.096095 &  122.733445 &    0.164955 \\
    16&  125.801781 &    0.157773 &   88.981748 &    0.093776 &  128.870117 &    0.159648 \\
    17&  135.006790 &    0.144317 &   96.652588 &    0.090859 &   76.708403 &    0.140725 \\
    18&  115.062605 &    0.143714 &   84.379244 &    0.090526 &   73.640067 &    0.120937 \\
    19&   75.174235 &    0.138975 &  111.994269 &    0.066164 &   85.913412 &    0.116579 \\
    20&  128.870117 &    0.113322 &   95.118420 &    0.062765 &   95.118420 &    0.105852 \\                                                                                        
 \hline\hline
    1&  131.938454 &    0.817488 &  119.665109 &    1.000000 &  131.938454 &    0.918017 \\
     2&  119.665109 &    0.736141 &  127.335949 &    0.760479 &  133.472622 &    0.801207 \\
     3&  118.130941 &    0.689816 &  124.267613 &    0.750421 &  121.199277 &    0.707666 \\
     4&  130.404285 &    0.663555 &  118.130941 &    0.729137 &  118.130941 &    0.648115 \\
     5&  133.472622 &    0.647121 &  131.938454 &    0.702722 &  130.404285 &    0.630494 \\
     6&  124.267613 &    0.600851 &  130.404285 &    0.695045 &  116.596773 &    0.629587 \\
     7&  127.335949 &    0.539153 &  125.801781 &    0.493815 &  122.733445 &    0.570149 \\
     8&  121.199277 &    0.510406 &  133.472622 &    0.442264 &  115.062605 &    0.445213 \\
     9&  122.733445 &    0.455214 &  135.006790 &    0.441181 &  124.267613 &    0.398636 \\
    10&  116.596773 &    0.449717 &  128.870117 &    0.384844 &  119.665109 &    0.289495 \\
    11&  125.801781 &    0.356472 &  122.733445 &    0.298950 &   78.242571 &    0.257532 \\
    12&  115.062605 &    0.327240 &   79.776739 &    0.215070 &   76.708403 &    0.225199 \\
    13&  135.006790 &    0.312036 &   78.242571 &    0.200541 &  141.143462 &    0.178793 \\
    14&  128.870117 &    0.276401 &  142.677630 &    0.185873 &  111.994269 &    0.175189 \\
    15&   78.242571 &    0.230802 &   82.845075 &    0.159508 &  113.528437 &    0.162036 \\
    16&   79.776739 &    0.185546 &  136.540958 &    0.155262 &   79.776739 &    0.150332 \\
    17&   76.708403 &    0.171929 &  113.528437 &    0.155105 &  142.677630 &    0.148415 \\
    18&  142.677630 &    0.168190 &   81.310907 &    0.148642 &  139.609294 &    0.130068 \\
    19&  113.528437 &    0.158608 &  144.211798 &    0.144056 &  110.460101 &    0.126255 \\
    20&  141.143462 &    0.132717 &  121.199277 &    0.142257 &  147.280134 &    0.122198 \\

 \hline\hline 
    1&  130.404285 &    0.595491 &  130.404285 &    0.602931 &  128.870117 &    0.611627 \\
     2&  131.938454 &    0.461723 &  131.938454 &    0.595086 &  130.404285 &    0.587957 \\
     3&  128.870117 &    0.436930 &  139.609294 &    0.558446 &  133.472622 &    0.420099 \\
     4&  136.540958 &    0.428092 &  136.540958 &    0.556658 &  135.006790 &    0.410902 \\
     5&  139.609294 &    0.394928 &  138.075126 &    0.453658 &  147.280134 &    0.392487 \\
     6&  147.280134 &    0.393585 &  147.280134 &    0.394679 &  127.335949 &    0.314830 \\
     7&  138.075126 &    0.321138 &  148.814302 &    0.375144 &  145.745966 &    0.313433 \\
     8&  135.006790 &    0.310646 &  141.143462 &    0.310228 &  144.211798 &    0.286731 \\
     9&  127.335949 &    0.307596 &  125.801781 &    0.306661 &  142.677630 &    0.277177 \\
    10&  133.472622 &    0.305220 &  127.335949 &    0.300189 &  131.938454 &    0.268791 \\
    11&  148.814302 &    0.285385 &  142.677630 &    0.285142 &  108.925933 &    0.257125 \\
    12&  142.677630 &    0.281188 &  107.391764 &    0.245189 &  136.540958 &    0.238031 \\
    13&  145.745966 &    0.271008 &  145.745966 &    0.220568 &  118.130941 &    0.185580 \\
    14&  141.143462 &    0.220437 &  105.857596 &    0.165661 &  148.814302 &    0.148852 \\
    15&  125.801781 &    0.218882 &  116.596773 &    0.165479 &  150.348470 &    0.135856 \\
    16&  144.211798 &    0.208107 &  135.006790 &    0.155439 &  107.391764 &    0.133676 \\
    17&  107.391764 &    0.197467 &  110.460101 &    0.149926 &  104.323428 &    0.129141 \\
    18&  108.925933 &    0.184959 &  121.199277 &    0.147039 &  116.596773 &    0.111140 \\
    19&  118.130941 &    0.143629 &  122.733445 &    0.138278 &  105.857596 &    0.108521 \\
    20&  116.596773 &    0.140953 &  113.528437 &    0.134046 &  102.789260 &    0.099456 \\

  \hline  
  \hline
 \end{tabular}
\end{center}
\end{table*}

\begin{figure}
\includegraphics[angle=0, width=80mm]{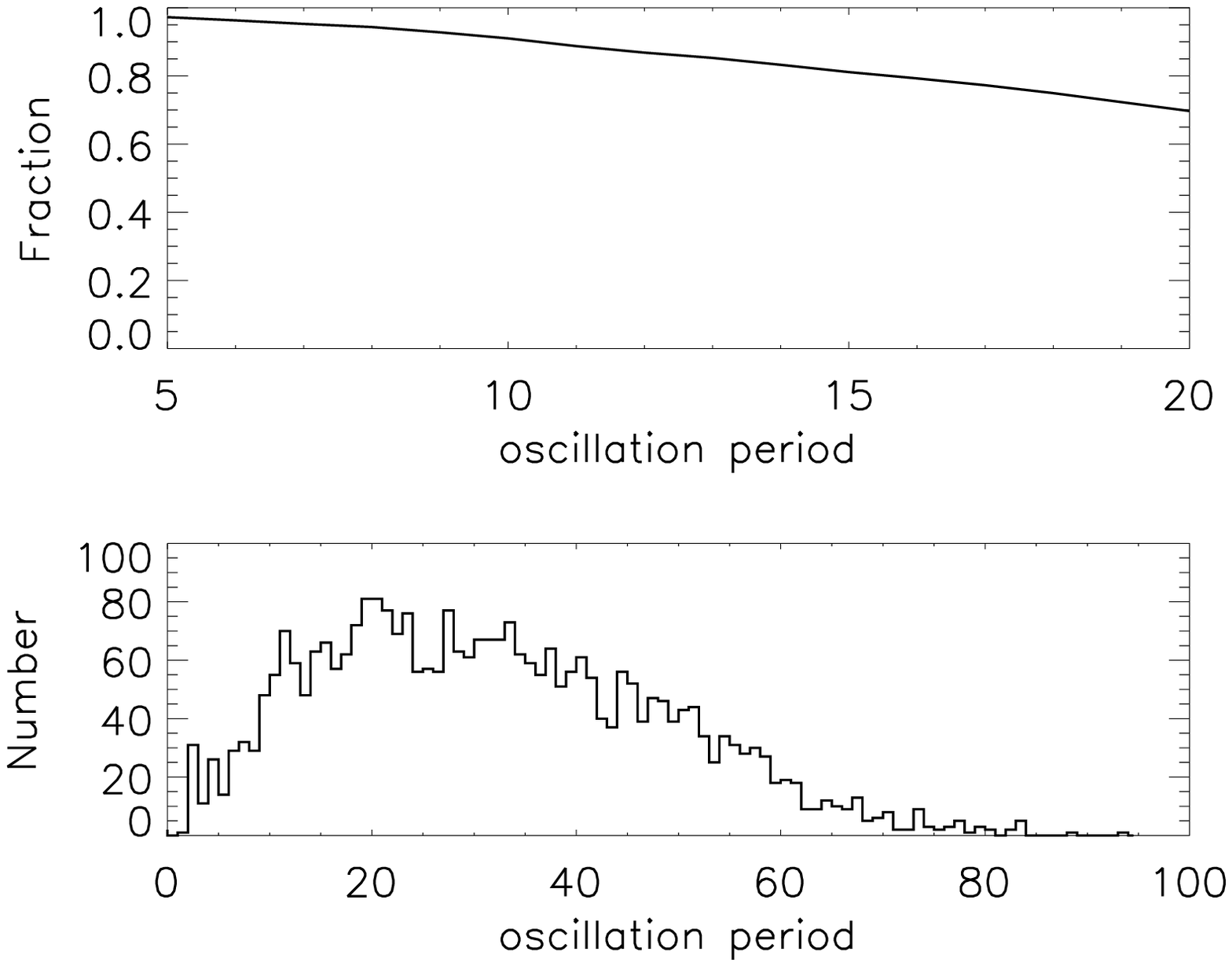}
\caption{Top: Fraction of the orbits having completed at least a given number of oscillation periods during the the time we follow our simulation, as a function of this number of oscillation periods.  Bottom: Histogram of the oscillation periods.}
\label{fig:peri}
\end{figure}

\begin{figure}
\includegraphics[angle=0, width=80mm]{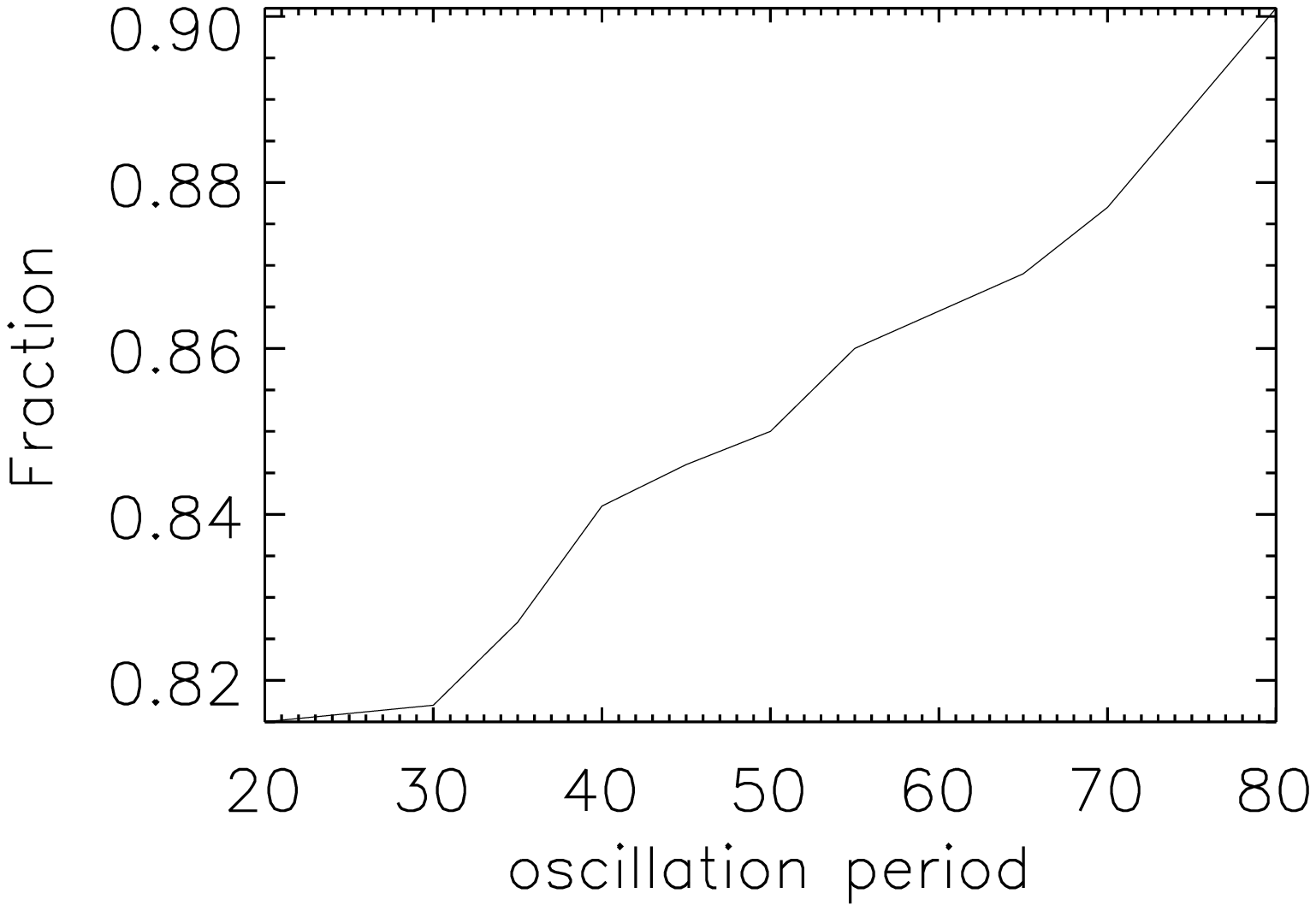}
\caption{Dependence of the fraction of the orbits with the same main frequencies in both NAFF and CA methods on the orbit oscillation periods.}
\label{fig:frac_peri}
\end{figure}

As shown in \cite{2010MNRAS.403..525V}, the accuracy of the frequency analysis decreases significantly when orbits were integrated for less than 20 oscillation periods, therefore, it is interesting to compare the main frequencies from the NAFF method with those from the CA method for orbits with more than 20 oscillation periods. In the top panel of Figure~\ref{fig:peri}, we show the fraction of our orbits with fixed oscillation periods. We find that $70\%$ orbits have more than 20 oscillation periods. In the bottom panel of Figure~\ref{fig:peri}, we present the histogram of the oscillation periods. It is noted that the distribution peaks  around 20. Therefore, the output interval for most orbits in our sample is reasonable for the frequency analysis. Figure~\ref{fig:frac_peri} shows that the fraction of orbits having three frequencies in agreement ($\delta f_{2,0}=0.1$) from the two methods increases strongly with the number of orbit oscillation periods. 
We find that, for $90\%$ of the orbits having undergone at least 80 oscillation periods, the three frequencies as calculated by the CA and NAFF methods agree. 
On the other hand,  for orbits which have less than 5 oscillation periods, only $39\%$ have three frequencies in agreement between two methods.


\section{fraction of regular orbits as obtained from NAFF and CA}

\begin{figure}
\includegraphics[angle=0, width=80mm]{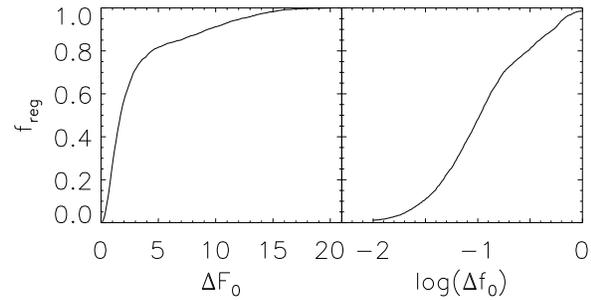}
\caption{Dependence of the fraction of regular orbits on the critical frequency drift parameter from NAFF. $Left$: For the absolute frequency drift. $Right$:  For the relative frequency drift.}
\label{fig:frac_reg_naff}
\end{figure}

Once we have the main frequencies, NAFF  classifies orbits as regular
or chaotic using the frequency drift. CA  classifies the orbits by finding the number of the fundamental frequencies. 
Figure ~\ref{fig:frac_reg_naff} shows the dependence of the fraction of regular orbits on the absolute critical frequency drift parameter (Left panel) and the relative critical frequency drift 
parameter (Right panel) from the NAFF method.  It is seen that the
fraction of regular orbits strongly depends on the critical frequency
drift value, but it is difficult to give a reasonable choice.

\begin{figure}
\includegraphics[angle=0, width=80mm]{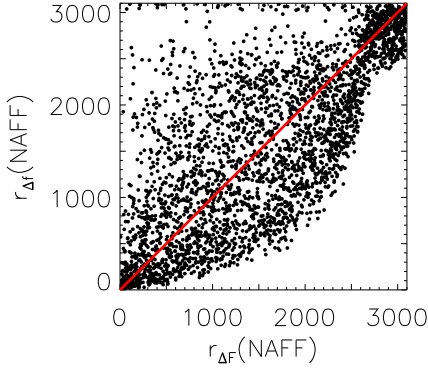}
\caption{Comparison of the ranking of the orbits in regularity by
  $\Delta F $ and  $\log(\Delta f)$. The red solid line represents the
  equality of the two rankings from $r_{\Delta F}$ and $r_{\Delta f}$.}
\label{fig:rank_naff}
\end{figure}

In order to compare the ranking of the various orbits in regularity by two kinds of the critical value in the NAFF method, we rank all orbits as a function of their $\Delta F$ values. The ranking is defined as $r_{\Delta F}$. 
The most regular orbit will have $r_{\Delta F}=1$ and the most chaotic one $r_{\Delta F}=3094$. We then rank 3094 orbits as function of their $\log(\Delta f)$ values, which is called $r_{\Delta f}$. Again the most regular orbit will have $r_{\Delta f}=1$ and the most chaotic one $r_{\Delta f}=3094$.
Figure~\ref{fig:rank_naff} shows the comparison of  the ranking of the
orbits in regularity by $\Delta F $ and  $\log(\Delta f)$. We find that only 
727 orbits have  similar
rankings in $r_{\Delta F}$ and $r_{\Delta f}$ ($|r_{\Delta f}/r_{\Delta F} -1|\le0.1$ ). For most orbits, however, there is a large dispersion between $r_{\Delta F}$ and $r_{\Delta f}$.

\begin{figure}
\includegraphics[angle=0, width=80mm]{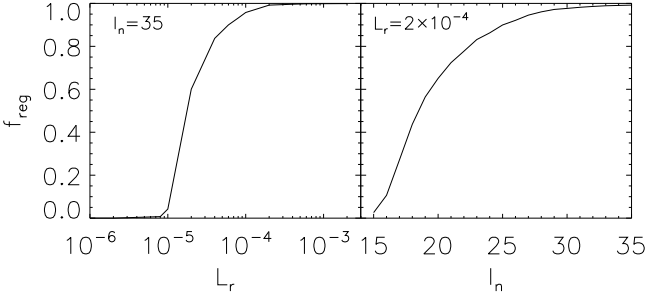}
\caption{Dependence of the fraction of regular orbits on the parameters $L_{\rm r}$ (left panel) and $I_{\rm n}$ (right panel) in CA. 
For the left and right panels, the parameters $I_{\rm n}$ and $L_{\rm r}$ are fixed at 35, and $2\times 10^{-4}$, respectively.} 
\label{fig:frac_reg_CA}
\end{figure}

In CA, an orbit is classified by the number of the fundamental
frequencies. An orbit is irregular if it has more than three
independent fundamental frequencies, otherwise it is regular. In this
method, there are two important parameters $L_{\rm r}$ and
$I_{\rm n}$. The former is used to determine whether the two main frequencies are the same, and whether the main frequencies are at resonance, while
the latter one  is the maximum integer for searching linear independence of the fundamental frequencies (see Sect. 4). 
Although the main frequencies weakly depends on the parameter $L_{\rm r}$, the number of fundamental frequencies strongly depends on it,
thus affecting  the fraction of regular orbits. In the left panel of
Figure ~\ref{fig:frac_reg_CA}, this fraction increases with the
increasing value of $L_{\rm r}$. If $L_{\rm r}$ is larger than
$10^{-4}$, most orbits are regular. In the right panel of Figure ~\ref{fig:frac_reg_CA}, we show the dependence of the fraction of regular orbits on the parameter $I_{\rm n}$. 
It is noted that most orbits are regular if $I_{\rm n}$ is larger than 25.        
$I_{\rm n}$=35 is usually chosen in order to classify correctly a large set
of orbits coming from selected known analytic potentials (Carpintero, private communication).  
In the bar system, nearly all orbits are regular if $I_{\rm n}$=35.

\begin{figure}
\includegraphics[angle=0, width=80mm]{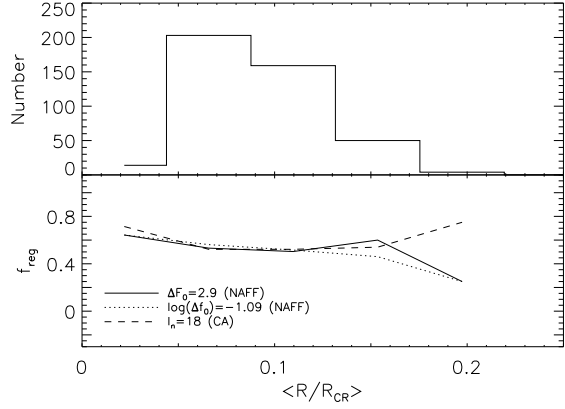}
\caption{Top: Number distribution of the orbits with more than 50 oscillation periods. Bottom: Fraction of regular orbits along the average radius of the orbits for orbits with more than 50 oscillation periods. The solid, dotted and dashed lines represent the results from the 
absolute critical value in NAFF, relative critical value in NAFF  and
CA, respectively. The different lines represent the results with
the different parameters in two methods.  
The region with $\langle R/R_{\rm CR}\rangle$  beyond 0.2 is ignored
because there are no orbits with more than 50 oscillation periods.}    
\label{fig:frac_reg_radii}
\end{figure}

The upper panel of Figure ~\ref{fig:frac_reg_radii} shows the number of orbits with more than 50 oscillations.   
It is a decreasing function of radius, as expected, because  the inner orbits have, on average, shorter orbital periods than the outer ones. All orbits with more than 50 oscillations have the average radius $\langle R/R_{\rm CR}\rangle$ 
smaller than 0.2. The lower panel of Figure~\ref{fig:frac_reg_radii} shows the
correlation between the fraction of regular orbits and the average orbit radius for orbits with more than 50 oscillations for the two methods with different parameters.
For $\Delta F_0=2.9$
or $\log(\Delta f_0)=-1.09$ in NAFF,  and $I_{\rm n}=18$ in CA, we find in general very good agreement between all methods, with a fraction of regular orbits around $53\%$.

It seems that
the regular fraction along the radius from NAFF with  $\log(\Delta
f_0)=-1.09$ is consistent with that from CA with $I_{\rm n}=18$ if
$\langle R/R_{\rm CR}\rangle$  is smaller than 0.18. If, however, we compare the
orbit types from the two methods one by one, we find a small decrease, so that only $47\%$ of the orbits
have the same type in both NAFF and CA methods. When we compare the two NAFF methods, we find that a very large fraction, $95\%$,
the same types when we use $\Delta F_0=2.9$ and $\log(\Delta f_0)=-1.09$.
in NAFF. 

If we take $\log (\Delta f_0)=-1.09$ and
consider the frequency drift parameter only in the x, y, and z
components, respectively, then the regular orbit fractions are
$56.2\%$,  $52.2\%$ and $72.8\%$.  If we consider the frequency drift
parameter in two components, x and y, x and z, y and z, then the
regular orbit fractions with  $\log (\Delta f_0)=-1.09$ are $43.2\%$,
$50.7\%$ and $46.7\%$, respectively. Therefore, the z component is
most regular, while  the y component is most chaotic in the bar
system. This is similar to the fact that the intermediate tube orbits
are unstable in the triaxial system
~\citep[e.g.][]{1996ApJ...460..136M, 2008gady.book.....B}.


         
\section{Discussion}
\begin{figure}
\includegraphics[angle=0, width=80mm]{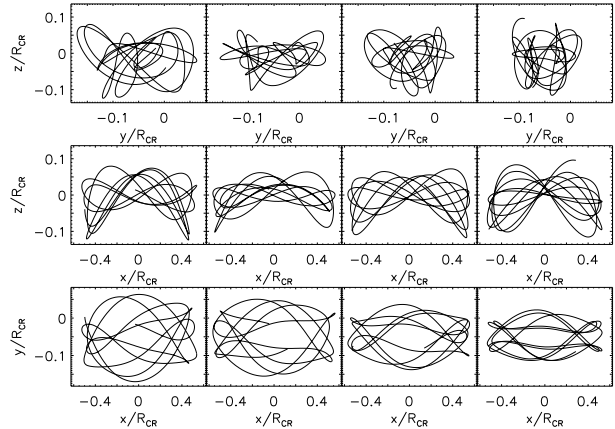}
\caption{Same as the left panel of Figure ~\ref{fig:orbit2745}, but for orbit 160, a orbit looks regular in the x-y and x-z planes, but chaotic in the y-z plane.} 
\label{fig:orbit160}
\end{figure}

It  seems difficult to give definite values of  $\Delta F_0$ and
$\log(\Delta f_0)$ in NAFF,  and $L_{\rm r}$ and $I_{\rm n}$ in CA to
classify orbits, but we can attempt to do this by selecting some
likely regular and chaotic orbits. We use 196 orbits which have the
same main frequencies from the NAFF and CA methods, and have a small
frequency drift $\Delta F<0.5$. We find that when we choose
$I_{\rm n}\ge30$ and $L_{\rm r}=2\times 10^{-4}$, most of these orbits
are regular in the CA method.  Even so, a few of these orbits, are
still irregular when we take $I_{\rm n}=30$. For example, as shown in
Figure~\ref{fig:orbit160},  orbit 160 is a regular orbit in NAFF, but
we find there are some chaotic property  in the y-z plane and this
could explain why CA classifies it irregular. 

\begin{figure}
\includegraphics[angle=0, width=80mm]{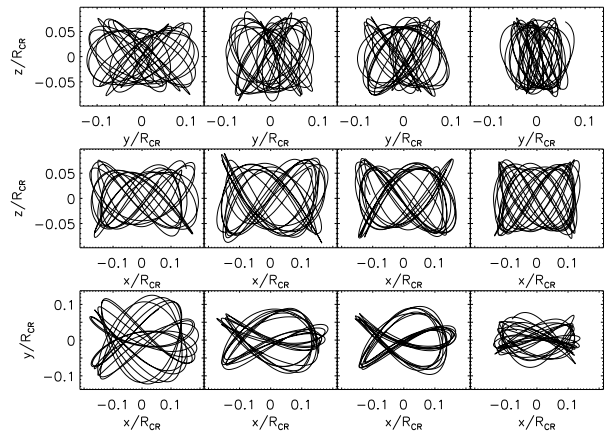}
\caption{Same as the left panel of Figure ~\ref{fig:orbit2745}, but for orbit 865, a orbit being regular in each interval, but the shape changing with time.} 
\label{fig:orbit865}
\end{figure}

Next we select 40 orbits which have the same main frequencies from the
NAFF and CA methods, but with large frequency drift  $\Delta F>9.3$.
Most of them are irregular when we take $I_{\rm n}=16$, but orbits such as
orbit  865, are still regular in the CA method.  From
Figure~\ref{fig:orbit865}, usually this orbit is regular  in each
interval, but the shape changes with time. Since the three main
frequencies are independent and no extra fundamental frequencies are found, the CA method classifies
it as a regular box orbit. On the other hand, if we use the
frequency drift method to classify it, it will be classified as
irregular. This frequency drift, however, could perhaps be due to the slight
potential changes with time and may not necessarily be due to
the fact that the orbit is irregular.

There is a further point related to the evolution of the potential. 
Namely, we find that the smallest $\Delta F$
  among the full 3094 orbits is 0.016, while the resolution of the
  main frequencies is usually $10^{-4}-10^{-3}$. Thus, even the
  smallest $\Delta F$ is still larger than the frequency resolution.
If we take  $\Delta F_0=0.01$, then every orbit is chaotic in NAFF; a
small $I_{\rm n}$ in CA with $L_{\rm r}=2\times 10^{-4}$ can give
similar results, so we can say both methods are in  good agreement , but
this is only an extreme case. Compared with the CA method,  the
results of NAFF only weakly depend on $L_{\rm r}$ and $L_{\rm
  max}$. For the parameter $L_{\rm max}$, if we do not take into account 
CPU time limits, we can make it as large as possible. Also the $I_{\rm n}$
value may have to be chosen differently for different potentials in the CA method. The
advantage of the CA method is that it can give  independent
fundamental frequencies of orbits, which can yield more detailed
information about regular orbits.  

\section{Summary and conclusions}

Individual particle orbits are the backbone of any structure. It is
thus important for understanding the formation and evolution of this structure to know whether the orbits that constitute it are chaotic or regular
and, in the latter case, what family they are associated with. Bars,
in particular, are a favourite field for such tests and thus many
studies have addressed the orbital structure in bars. Most of them,
however, use an analytic potential and are thus not very realistic
\citep[see e.g.][ for a review]{2016ASSL..418..391A}. A further
disadvantage of such 
studies is that it is not trivial to choose the initial conditions for
the orbits and the result can depend critically on this
choice. Instead, we used here orbits taken directly from the
simulation. This means that they have very realistic potentials, but
at the expense of some noise and, particularly, some evolution of the
potential. 

As a first step towards understanding the orbital structure in bars, we
compare in this paper two methods, the NAFF method of \cite{1990Icar...88..266L}     
and the method of \cite{1998MNRAS.298....1C}. 

We show how the main frequencies depend on
the maximum extracted line number $L_{\rm max}$ and on the parameter
to distinguish two main frequencies  $L_{\rm r}$. We find that only a small number $(<0.1\%)$ of the main frequencies 
in NAFF have been changed when using different values of $L_{\rm max}$, while about $6\%$ of the main frequencies have been changed in CA.
If we change $L_{\rm r}$ from $2\times10^{-4}$ to $2\times 10^{-3}$, then around $6\%$ and $1.5\%$ of the orbits have a different main frequency in the NAFF 
and CA methods, respectively.

We find that, at  least for our case, the main frequencies calculated by the
two methods are in good agreement provided we use the same
definitions and values for $L_{\rm max}$ and  $L_{\rm r}$: for $80\%$
of the orbits the differences between the results of the two methods
are less than $5\%$ for all three main frequencies.  We also find that there are two clear peaks  in the histogram of the ratios of the three main frequencies in both methods. 
The highest peak is 1:1, and the second highest is 2:3 for the face-on view $(x,y)$.  The two edge-on views, $(y,z)$ and $(x,z)$ also have two clear peaks, one at 4:5 and the other at 4:7.

We find that the fraction of the regular orbits strongly depends on two parameters $L_{\rm r}$ and $I_{\rm n}$ in the CA method. 
The former is used to determinate whether the two frequencies are the same and whether there are resonances among the main frequencies. 
The fraction of the regular orbits increases with increasing  $L_{\rm r}$ or $I_{\rm n}$.
In the NAFF methods, the fraction of the regular orbits strongly depends on the critical frequency drift parameter.
The regular fraction is increased with increasing this parameter.  However, it is difficult to give certain values of these parameters in both methods.  
The fact that there is no abrupt change from chaotic to regular reflects the fact that there is stickiness and confined chaos. 
We also find that, for a given particle, in general the projection of its motion along the bar minor axis is more regular than the other two projections, 
while the projection along the intermediate axis is the least regular.  

Increasing the number of particles in the simulation will decrease the
noise. In a future paper we plan to use a simulation with a
considerably larger number of particles, to determine how noise may influence the
results.

\section*{Acknowledgements}
We thank the referee for comments and suggestions that improved the paper. We acknowledge helpful discussions with Jacques Laskar, Hongsheng Zhao and helpful email exchanges with Daniel Carpintero and Monica Valluri. We also thank Jeans-Charles Lambert 
for his help with software and computing.  We also thank Daniel Carpintero for making available to us the latest version of his code. 
This work is supported by the Strategic Priority Research Program ``The Emergence of Cosmological Structures" 
of the Chinese Academy of Sciences (Grant No. XDB09000000). We also acknowledge the support by the 973 Program (No. 2014CB845700), the National Science
Foundation of China (Grant No. 11390372, 11333003, Y011061001 and Y122071001), and the Sino-French ``Lia-Origins" project. 

This work was started during the 2011 workshop on the Galactic bulge
and bar in the Aspen Center for Physics. We thank the hospitality of
the Aspen Center for Physics, which is supported by the NSF grant
1066293. The simulation used here was run using the HPC resources of
[TGCC/CINES/IDRIS] under the allocations 2013-[x2013047098] and
2014-[x2014047098] made
by GENCI. Most of the remaining computing was performed on the
supercomputer ``laohu" at 
the High Performance Computing Center at the National Astronomical
Observatories of China, funded by the Ministry of Finance under the
grant ZDYZ2008-2.

\bibliographystyle{mn2e}
\bibliography{ms}

\appendix

\label{lastpage}

\clearpage
\end{document}